\documentclass[aps,prb,twocolumn,superscriptaddress,10pt]{revtex4-1}
\usepackage[utf8]{inputenc}
\usepackage{libertine}
\usepackage[libertine]{newtxmath}
\usepackage{mathtools,graphicx,microtype,bm}
\usepackage[cal=boondoxo,frak=boondox]{mathalfa}
\usepackage[colorlinks=true,allcolors=blue]{hyperref}

\begin{document}
\title{Spin Hall effect in two-dimensional materials with inverted bands and Mexican-hat dispersion}

\author{Bagun S.\ Shchamkhalova}
\author{Vladimir A.\ Sablikov}
\email[E-mail:]{s.bagun@gmail.com} 
\affiliation{Kotelnikov Institute of Radio Engineering and Electronics, Fryazino Branch, Russian Academy of Sciences, Fryazino, Moscow District, 141190, Russia}

\begin{abstract}
We study the spin Hall effect in two-dimensional topological insulators with "Mexican hat" dispersion and a ring-shaped Fermi surface which are formed due to the band inversion.
Electron transitions between different isoenergetic contours and the quantum metric of band states play an important role in the transport properties of such materials, since they largely determine the spatial distribution of the electron charges screening the impurity potential and the scattering probability [Phys.B, 719, 417942 (2025)]. Here we study a spin-dependent skew scattering, which is enabled by the second-order scattering processes, and show that the extrinsic spin-Hall current (SHC) can significantly exceed the intrinsic SHC arising from the Berry curvature. Furthermore, due to Mexican-hat dispersion, the SHC exhibits a very unusual dependence on the Fermi energy ($E_F$). The extrinsic SHC reaches a maximum at some $E_F$, then decreases with increasing $E_F$ and can even change a sign. This complicated behavior reflects an interplay of energy dependencies of such important factors as probabilities of inter- and intra-contour transitions, as well as different electron velocities in two contours.
\end{abstract}

\maketitle

\section{Introduction}\label{S_1}
The spin Hall effect (SHE) in topological insulators (TIs)~\cite{Sinova,Sinitsyn_2008,RevModPhys.95.011002} has recently become one of the most actively studied topics in condensed matter physics. The presence of spin currents in the surface and edge states of TIs opens up broad prospects for their use in various systems for highly efficient conversion of charge currents into spin currents, accumulation of spin density, and generation of spin-orbit torque (SOT) in TI/magnetic systems~\cite{PhysRevLett.124.066401,PhysRevLett.134.056301,PhysRevB.106.024405}. Wide range of applications of magnetic devices, such as logic-in memory, magnetic random-access memory and spin torques that flip magnetic bits, requires an efficient method for controlling the local magnetization~\cite{Manchon1,588c-z8cy}. Higher charge-to-spin conversion efficiency is crucial for the operation of low-power SOT devices, but the mechanisms of charge-to-spin conversion and SOT formation are still poorly understood. In TIs, spin current and torque are generated both by the intrinsic SHE, caused by the topology of quantum states, and by the extrinsic SHE, caused by the spin-dependent impurity scattering. Their contributions to the total SHE are not well separated yet. 

This paper aims to study the SHE in a realistic model of a TI, consistently taking into account both the topological properties and the interaction of electrons with impurities, including the screening of the scattering potential which also depends on topology of band states.

Since the intrinsic topological SHE in topological insulators is studied quite well~\cite{PhysRevLett.96.106802,Yu2025}, we focus mainly on the extrinsic SHE, the role of which in realistic conditions is not yet well understood though it is known that extrinsic SHE can significantly contribute into the total SHE~\cite{Sinova,Sinitsyn_2008}. The magnitude of the extrinsic spin-Hall effect is determined by the spin Hall current generated as a result of asymmetric spin-dependent scattering of electrons by charged impurities in the bulk. To elucidate the physics of spin current generation and estimate its magnitude, we will study the spin current in the bulk of the sample far from the boundaries. Boundary effects such as edge currents, spin density accumulation~\cite{Hirsch,Manchon}, and boundary conditions at the Hall contacts are ignored at this stage.

Recently, much attention has been paid to atomically thin materials with Mexican hat dispersion (MHD) in the valence or conduction band which is realized in many materials such as
HgTe/CdHgTe quantum wells~\cite{Krishtopenko}, biased graphene bilayer~\cite{Stauber}, the semiconducting III-VI monochalcogenides, GaS, GaSe, InS, and InSe~\cite{10.1063/1.4928559}, rhombohedral $\alpha$-In$_2$Se$_3$~\cite{kremer2025}, double~\cite{PhysRevB.95.045116, PhysRevLett.119.056803} and triple~\cite{meyer2025} InAs/GaSb quantum wells, transition-metal halogenides~\cite{10.1063/5.0237686}, Sn-doped Bi$_{1.1}$Sb$_{0.9}$Te$_2$S~\cite{PhysRevB.101.121115}. Some aspects of charge transport properties due to the density of states in MHD materials were recently discussed in Ref.\,\onlinecite{Rukelj}

Of particular note are the strained layer InAs/In$_{x}$Ga$_{1-x}$Sb quantum well structures~\cite{Zhang}, which not only have a wide inverted band gap, but also MHD in both the valence and conduction bands. Transport measurements  confirmed the preservation of topologically protected edge transport in the presence of enhanced bulk strain  within the In$_{0.5}$Ga$_{0.5}$Sb quntum wells. 

In this paper, we study the skew scattering in a 2D TI with a MHD, which is expected to result in a stronger skew scattering effect.
In this case the basis quantum states of electrons have non-trivial quantum-geometric properties, both the quantum metric and the Berry phase, which turn out to be significantly larger for systems with MHD~\cite{Sablikov1}. 

The important features of MHD are the Van Hove singularity of the density of states (DOS) at the MHD bottom and a presence of the two Fermi contours. In the energy range between the bottom and the top of the MHD, there are two Fermi contours and when studying the scattering of electrons in this energy range one needs to take into account electron transitions both between the two Fermi contours and within each contour.  Due to the combined effect of these two factors, the polarization and the screening acquire unique properties in this materials~\cite{Sablikov1, PhysRevB.99.085409,Shchamkhalova}. 

Another feature is related to the effective mass of quasiparticles. On the low-wave vector branch of the MHD, the effective mass changes sign with increasing energy from positive near the MHD bottom to negative near the top. This obviously affects the  scattering of electrons and their density distribution around the external charge and, consequently, the screened potential, since quasiparticles with negative mass are attracted to the negatively charged center. Due to this feature, not only quasi-bound states with energy above the MHD top are formed~\cite{SABLIKOV2023115492,SABLIKOV2023129006}, but skewness of the scattering changes the sign.  

We calculated both the extrinsic and intrinsic SHCs and show that the extrinsic SHC can significantly exceed the intrinsic SHC at a low temperature and the low $E_F$. This occurs when the extrinsic SHC is mainly caused by electron transitions near the inner Fermi contour. With the increase of the $E_F$ the extrinsic SHC exhibits very unusual dependence on the $E_F$ which appears due to the MHD. When the MHD shape is deep enough, the extrinsic SHC reaches a maximum and then decreases and even changes its sign. The effect of the SHC sign change disappears when the MHD shape becomes flatter. In any case the SHC dependence on $E_F$ reflects the interplay of the energy dependencies of such important factors as the inter- and intra-contour transitions of electrons and the velocities of electrons in two contours.

In MHD materials the electron momentum is not a single-valued function of an energy and therefore the process of elastic relaxation of the distribution function of charge carriers takes place through three channels, including transitions inter two isoenergetic Fermi-contours along with intra-contour ones. When voltage is applied, the distribution functions for the two dispersion branches become different and, in addition, asymmetric with respect to the scattering angle. The Boltzmann equation in the case of two dispersion branches can be solved~\cite{Shchamkhalova} reducing it to a system of linear equations. Here we have developed this approach to calculate asymmetric distribution functions for each of the two dispersion branches: the Boltzmann equation is reduced to a system of four equations to determine the SHC. 

The calculations of the screened impurity potential are performed in the random phase approximation (RPA). The scattering matrix is found within the Born approximation by expanding the scattering amplitude to the third order in the scattering potential, where spin-dependent scattering appears.
All specific calculations are carried out within the frame of the Bernevig-Hughes-Chang (BHZ)~\cite{BHZ} model which is rather universal model of TIs. Within this model the MHD arises due to the inversion of the electron and hole bands when their hybridization is not too strong. 

In Sec.~\ref{S_potential}, we describe the model. In Sec.\ref{wave_function} the scattering is considered. The electron distribution function and conductivity are studied in Sec.~\ref{Boltzmann}. In Sec.~\ref{S_conclusion} we discuss the main results and give conclusions.

\section{Model Hamiltonian and screened impurity potential}\label{S_potential}

In the spatial inversion symmetric BHZ model~\cite{BHZ}, where the spin component perpendicular to the layer is a good quantum number, the total Hamiltonian splits into two Hamiltonians, one for each spin orientation, $s=\uparrow, \downarrow$. The spin-up Hamiltonian is~\cite{BHZ}:
\begin{equation}\label{eq.Hamiltonian}
    H_{\uparrow}= -Dk^2 +
    \begin{pmatrix}
        M-B \hat{k}^2 & A (\hat{k}_x+i\hat{k}_y)\\-
        A (\hat{k}_x-i\hat{k}_y) & -M+B \hat{k}^2\\
    \end{pmatrix},
\end{equation}
where the parameter $D$ describes the electron and hole bands asymmetry, $M$, $B$ and $A$ are standard parameters of the model~\cite{JPSJ.77.031007}.

We use dimensionless quantities. The values of the energy dimension are normalized to $|M|$, the distance is normalized to $\sqrt{|B/M|}$, the wave vector $k$ is normalized to $\sqrt{|M/B|}$. An important parameter of the model $a=A/\sqrt{|B M|}$ describes the hybridization of the electron and hole bands, $\delta=D/|B|$ describes an electron-hole asymmetry. The MHD is realized when $|a|<\sqrt{2}$. The dispersion relation is 
\begin{equation}\label{spectr}    
\varepsilon_\lambda(k)=-\delta k^2 +\lambda\varepsilon_k,
\end{equation}
where $\varepsilon_k=\sqrt{(1-k^2)^2+a^2k^2}$ and $\lambda=\pm 1$ is the band index ($\lambda=+1$ for conduction band and $\lambda=- 1$ for valence band).

Two isoenergetic contours with energy $\varepsilon$ are defined in a momentum space by $$k_1(\varepsilon) =\sqrt{(1 - a^2/2 + \varepsilon\delta - \Delta)/(1 - \delta^2)}$$ $$k_2(\varepsilon)=\sqrt{(1 - a^2/2 + \varepsilon\delta + \Delta)/(1 - \delta^2)},$$  where $$\Delta=\sqrt{a^2(a^2/4 - 1) + (\varepsilon + \delta)^2 -\varepsilon a^2\delta}.$$ 
At zero temperature, all momentum states located between two circles — the inner one with radius $k_{F1} = k_1(E_F)$ and the outer one with radius $k_{F2}= k_2(E_F)$ — are occupied.

\begin{figure}
    \centerline{\includegraphics[width=1\linewidth]{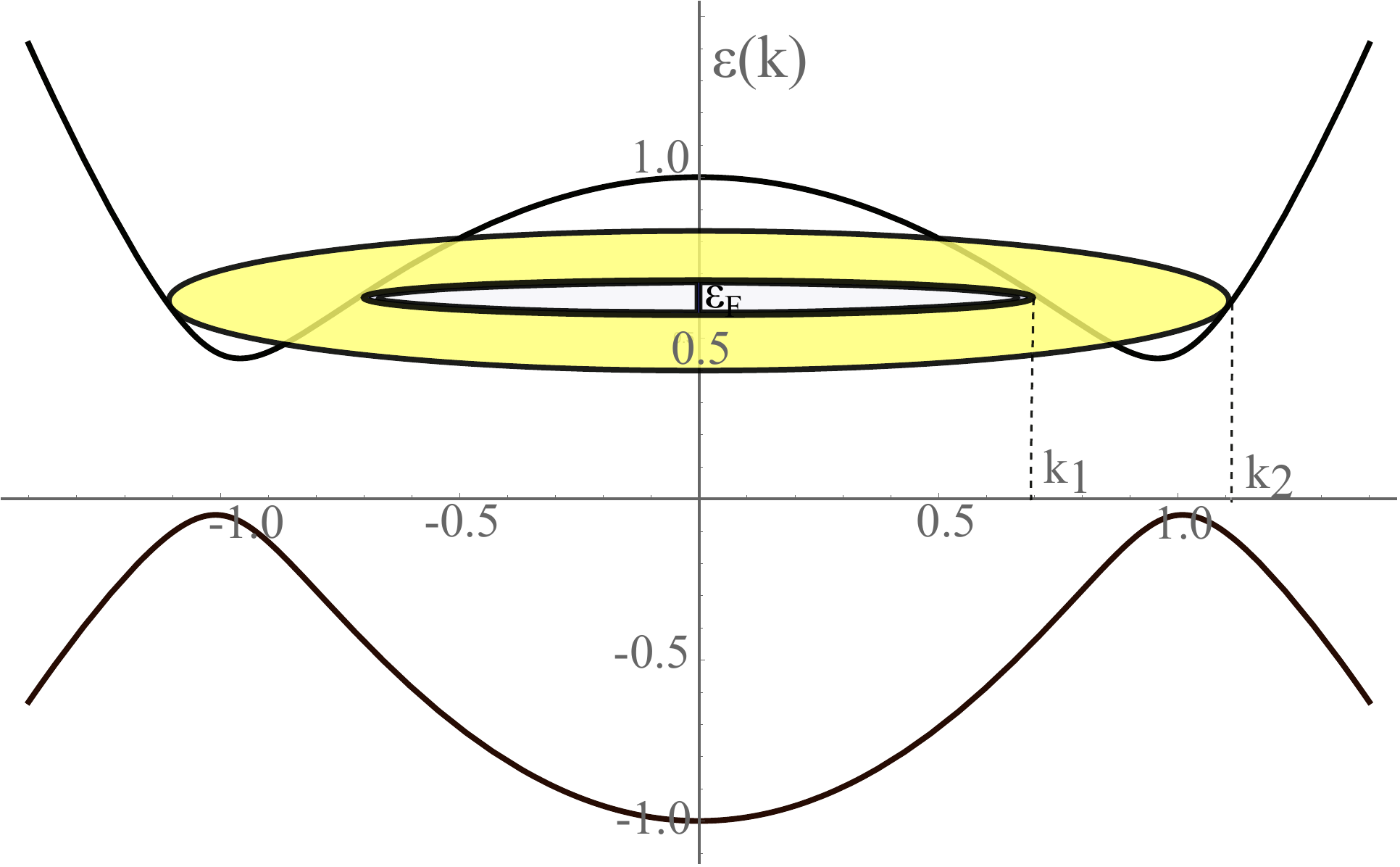}}
    \caption{MHD and ring-shaped Fermi surface for $a = 0.25$ and $\delta = -0.2$, conduction band minimum is at $E_0 =0.43$.}
\label{f1:fig1} 
\end{figure}

The energy dispersion in conduction band $\varepsilon_+( k)$ and the ring-shaped Fermi surface for the  $E_F = 0.6$  are shown in Fig.~\ref{f1:fig1} for parameters $a = 0.25$ and $\delta = -0.2$. The conduction band minimum equals  $E_0 =0.43$. All calculations below are done counting the $E_F$ from corresponding conduction band minimum. 

The spinor $u_{s, \lambda}(\bm{k})$ for spin-up states has the form
\begin{equation}\label{u-sp}
    u_{\uparrow, \lambda}( \bm{k}) = \frac{1}{\sqrt{1+\beta_{\lambda, k}^2}}
    \begin{pmatrix}
        1 \\ \beta_{\lambda, k} e^{-i\phi_k}  
    \end{pmatrix}\,,
\end{equation}
where 
\begin{equation}
    \beta_{\lambda, k}=\frac{a k}{\lambda\varepsilon_k-1+k^2}\,
\end{equation}
and $\phi_k$ is the polar angle of the vector $\bm{k}$.

The wave function of the band state with energy $\varepsilon_\lambda(k)$ is 
\begin{equation}
    |\uparrow\lambda, \bm{k}\rangle =\frac{1}{L}u_{\uparrow\lambda}(\bm{k})e^{i\bm{k r}}\,,
\end{equation}
with the wave vector $\bm{k}$ , $L$ is a normalization length. 

The Berry curvature of the state $|s,\lambda,\bm{k}\rangle$ is equal to
\begin{equation}\label{Berry}
    \bm{\Omega}_{s,\lambda}(\bm{k})=-s\lambda\frac{a^2(1+k^2)}{2\varepsilon_k^3}\bf{e}_z\,,
\end{equation}
where spin index $s=\pm 1$.

The velocity of charge carrier in  the state $|s,\lambda,\bm{k}\rangle$ equals
\begin{equation}\label{velocity}
\dot{\bm{r}}=\bm{v}_k+\frac{e}{\hbar}[\bm{E}\times\bm{\Omega}],
\end{equation}
where  $\bm{v}_k=(1/\hbar)(\partial{\varepsilon_{k}}/{\partial\bm{k}})$.

Within the RPA, the 2D Fourier transform of the screened impurity Coulomb potential equals~\cite{Sablikov1}: 
$$V(q)= \frac{\tilde{V}_q}{1-C_q\Pi(q,0)/q}=\frac{ZC_q}{q-C_q \Pi(q,0)},$$
where $\tilde{V}_q = ZC_q/q$ is 2D Fourier transform of the bare impurity Coulomb potential $V_0(r)=e^2Z/ \epsilon_0r$,  the impurity is supposed to be a point charge $Ze$ at position $\bm{r} = 0$, and $C_q=2\pi e^2/\epsilon_0\sqrt{|MB|}$ and $\Pi(\bm{q},0)$ is the Lindhard polarization function. 

Numerical calculations are performed for systems with parameters close to that of InAs/In$_{x}$Ga$_{1-x}$Sb quantum spin Hall insulators with strained layers. These structures have the advantage of having a sufficiently large spectral gap. In addition, their band dispersion can be controlled by the gate potential. Recent study of Zhang \textit{et.al.}~\cite{Zhang} of the strained-layer InAs/In$_{0.5}$Ga$_{0.5}$Sb within the eight-band Kane model shows results that are qualitatively consistent with the BHZ model with $a \approx 0.25$ and $\delta\approx - 0.2$.

\section{Scattered wave function}\label{wave_function}

The wave function of a scattered electron  are determined from
\begin{equation}
    (\widehat{H}_{\uparrow} +\widehat{V})|\Psi \rangle = E|\Psi \rangle,
\end{equation}
where $\widehat{V}(r)$ is the scattering potential calculated within the RPA framework taking into account the screening charge of electrons, 
\begin{equation}
   \widehat{V}(r)=\int\frac{L^2d^2q}{(2\pi)^2}V(q)e^{i\bm{q}\bm{r}}\,.
\end{equation}
The scattered wave function is presented in $\widehat{T}$-matrix formalism:
\begin{equation}
   |\Psi \rangle =|\uparrow\lambda, \bm{k}\rangle + \widehat{G}\widehat{V}|\Psi \rangle \equiv |\uparrow\lambda, \bm{k}\rangle +\widehat{G}\widehat{T}|\uparrow\lambda, \bm{k}\rangle\,, 
\end{equation}
where $\widehat{T}$ matrix is presented in terms of the Green function 
\begin{equation}
    \widehat{T}=\widehat{V} + \widehat{V}\widehat{G}\widehat{V} + ...\, 
\end{equation}
with 
\begin{equation}
    \widehat{G}(\uparrow\bm{r},\uparrow\bm{r}')=\sum_{\lambda}\int \frac{d^2k}{(2\pi)^2}\frac{|\uparrow\lambda, \bm{k}\rangle \langle\uparrow\lambda, \bm{k}| }{E-\varepsilon_\lambda(k)+i0}.
\end{equation}

The scattering matrix elements read:
\begin{widetext}
\begin{equation}
\begin{split}
 \langle\lambda', \bm{k}'|\widehat{V}|\lambda, \bm{k}\rangle = u^+_{\lambda}(\bm{k}) u_{\lambda'}(\bm{k'})\int\frac{d^2r}{ L^2}V(r)e^{i(\bm{k}'-\bm{k})\bm{r}} =
		\frac{1+\beta_{\lambda', \bm{k}'}\beta_{\lambda, \bm{k}}e^{i(\phi_{k'}-\phi_k)}}{\sqrt{(1+\beta_{\lambda', \bm{k}'}^2)(1+\beta_{\lambda, \bm{k}}^2)}}\frac{V_q}{L^2},
\end{split}
\end{equation}
%\end{widetext}
where $\bm{q}=|\bm{k}'-\bm{k}|=\sqrt{k^2+k'^2-2kk'\cos(\phi_{k'}-\phi_k)}$. 

Spin indexes are omitted in scattering matrix elements here and below because the scattering potential is independent of the spin. Here and below we present the results for the case of a completely occupied valence band and a low enough temperature, when only the elastic scattering in the conduction band is considered, so we omit below the band index $\lambda$.
\begin{equation}
     |\langle\bm{k}'|\widehat{T}|\bm{k}\rangle|^2 = |\langle\bm{k}'|\widehat{V}|\bm{k}\rangle|^2 + \langle\bm{k}'|\widehat{V}|\bm{k}\rangle\langle\bm{k}'|\widehat{V}\widehat{G}\widehat{V}|\bm{k}\rangle^* + c.c. + ...\, 
\end{equation}
$|\langle\bm{k}'|\widehat{V}|\bm{k}\rangle|^2 =|\langle\bm{k}|\widehat{V}|\bm{k}'\rangle|^2$ is an even function of the scattering angle. The second term which is $\propto V^3$ contains asymmetric in scattering angle part.
\begin{equation}
\label{Soch}
\begin{split}
     \langle\bm{k}'|\widehat{V}\widehat{G}\widehat{V}|\bm{k}\rangle = \int d^2k_1\frac{\langle\bm{k}'|\widehat{V}|\bm{k}_1\rangle\langle\bm{k}_1|\widehat{V}|\bm{k}\rangle}{E-\varepsilon_+(k_1)+i0}= \int dk_1k_1\frac{\Im_{k',k}(k_1)}{E-\varepsilon_(k_1)+i0} = P.V.\int dpp\frac{\Im_{k',k}(p)}{E-\varepsilon_+(p)}-i\pi\sum_{j=1}^2\frac{k_j\Im_{k',k}(k_j)}{(\frac{\partial{\varepsilon_{k}}}{{\partial k}})_{k_j}}
\end{split}
\end{equation}
where the index $j$ numbers two isoenergetic contours defined by the equation $\varepsilon(k_{i})=\varepsilon$ for MHD. 

We denote  $\Im_{k',k}(k_1)=\int d\phi_1\langle\bm{k}'|\widehat{V}|\bm{k}_1\rangle\langle\bm{k}_1|\widehat{V}|\bm{k}\rangle$:
\begin{equation}
\begin{split}
\Im_{k',k}(k_1) = \frac{1}{(2\pi)^4}\frac{1}{(1+\beta_{k_1}^2)\sqrt{(1+\beta_{k'}^2)(1+\beta_{k}^2)}}\int d\phi_1 V_{|\bm{k}_1-\bm{k}'|} V_{|\bm{k}-\bm{k}_1|}\\
 \times\left\{1+\beta_{k_1}\left[\beta_{k'}\exp[i(\phi'-\phi_1)] + \beta_{k}\exp[i(\phi_1-\phi)] \right]+\beta_{k_1}^2\beta_{k'}\beta_{k}\exp[i(\phi'-\phi)]\right\}.
\end{split}
\end{equation}
Only the last term of Eq.(\ref{Soch}) caused the asymmetric in scattering angle terms.   
\begin{equation}		
|\langle\bm{k}'|\widehat{T}|\bm{k}\rangle|^2\big|_{as}=	\frac{V_{|\bm{k}_1-\bm{k}'|}}{(2\pi)^2}\frac{1+\beta_{\lambda', \bm{k}'}\beta_{\lambda, \bm{k}}e^{i(\phi_{k'}-\phi_k)}}{\sqrt{(1+\beta_{\lambda', \bm{k}'}^2)(1+\beta_{\lambda, \bm{k}}^2)}}(-i\pi)\sum_{j=1}^2\frac{k_j\Im_{k',k}(k_j)}{|\frac{\partial{\varepsilon_{k}}}{{\partial k}}|_{k_j}} +c.c. 	
\end{equation}

Straightforward  calculations show that the asymmetric part of the electron scattering amplitude in the conduction band
\begin{equation}
    W(\bm{k},\bm{k}')=|\langle\bm{k}'|\widehat{T}|\bm{k}\rangle|^2\big|_{as}
\end{equation}
can be presented in the form:
%\begin{widetext}
\begin{equation}\label{Skew1}
\begin{split}
W(\bm{k},\bm{k}') = \frac{\sin\phi}{L^2}\sum_{i=1}^2\frac{\beta_{k'}\beta_{k}V_{|\bm{k}-\bm{k'}|}}{(1+\beta_{k'}^2)(1+\beta_{k}^2)(1+\beta_{k_i}^2)}\frac{k_i}{|v_i|}\left[(1-\beta_{k_i}^2)W_0(k,k',k_i,\phi) + \beta_{k'} W_1(k,k',k_i,\phi)+ \beta_{k} W_2(k,k',k_i,\phi)\right].
\end{split}		
\end{equation}
%\end{widetext}
We denote for brevity the scattering angle $\phi = \phi_{k'}-\phi_{k}$ and
$$W_0(k,k',k_i,\phi)=\int\frac{d\varphi}{(2\pi)^5}  V\left(\sqrt{k'^2+k_i^2-2k'k_i\cos\varphi}\right)V\left(\sqrt{k_i^2+k^2-2kk_i\cos(\varphi-\phi)}\right),$$
$$W_1(k,k',k_i,\phi)=\int \frac{d\varphi}{(2\pi)^5}  V\left(\sqrt{k'^2+k_i^2-2k'k_i\cos\varphi}\right)V\left(\sqrt{k_i^2+k^2-2kk_i\cos(\varphi-\phi)}\right)\cos\varphi,$$
$$W_2(k,k',k_i,\phi)=\int \frac{d\varphi}{(2\pi)^5}  V\left(\sqrt{k'^2+k_i^2-2k'k_i\cos(\varphi-\phi)}\right)V\left(\sqrt{k_i^2+k^2-2kk_i\cos\varphi}\right)\cos\varphi.$$
\end{widetext}

It is convenient to write $W(\bm{k},\bm{k}')$ in the form $W(\bm{k},\bm{k}') = \widetilde{W}(k,k',\phi)\sin\phi$, where $\widetilde{W}(k,k',\phi)$ is an even function of $\phi$. 

The matrix elements of intra- and inter-branch scattering in MHD systems significantly depends on the electron energy. The skew scattering intra-inner contour and inter-contours are the most intensive ones at a low $E_F$. With increasing energy the skew scattering matrix elements decrease rapidly. 

\section{Boltzmann equation and current}\label{Boltzmann}

Here we study the SHC generated by a weak electric field within the frame of the semiclassical approach~\cite{Sinitsyn_2008, Loss,CULCER2012860}. The problem of nonlinear SHE, which has attracted interest in recent years~\cite{PhysRevLett.134.056301,BANDYOPAD}, remains outside our scope.
The Boltzmann equation for homogeneous system in a stationary state can be written as:
\begin{equation}\label{Boltz}
\dot{\bm{k}} \frac{\partial f(\bm{k})}{\partial\bm{k}}=St[f],
\end{equation}
where $\dot{\bm{k}} =-(e/\hbar)\bm{E}$ and the collision term:
\begin{widetext}
\begin{equation}
St[f]=\frac{2\pi}{\hbar}N_i\int\frac{d^2k'}{(2\pi)^2}\left\{|\langle\bm{k}'|V|\bm{k}\rangle |^2+\widetilde{W}(k,k',\phi_{k'}-\phi_{k})\sin(\phi_{k'}-\phi_{k})\right\} [f(\bm{k'})-f(\bm{k})]\delta[\varepsilon(k)-\varepsilon(k')].
\end{equation}
Here, we suppose that the scattering Coulomb centers are not correlated and they scatter electrons independently, $N_i$ is the scattering impurity concentration. 

The collision integral can be simplified by explicitly tracking the two branches of the MHD~\cite{Shchamkhalova}, so that the Eq.~(\ref{Boltz}) becomes
\begin{equation}\label{Boltz1}
%\begin{split}
 -e\bm{v}_k\bm{E}\frac{\partial f}{\partial\varepsilon}=
\sum_{i=1}^2\frac{D_i(\varepsilon)}{2\pi\hbar^2}N_i\int d\phi_{k_i}\left\{|\langle\bm{k}_i|V|\bm{k}\rangle |^2+ \widetilde{W}(k,k_i,\phi_{k_i}-\phi_{k})\sin(\phi_{k_i}-\phi_{k})\right\} [f(\bm{k}_i)-f(\bm{k})],
%\end{split}		
\end{equation}
where $$D_i(\varepsilon)=\left.\frac{k_i(\varepsilon)}{\hbar |v_{k_i}(\varepsilon)|}={k_i(\varepsilon)}\right/\left|\frac{\partial{\varepsilon_{k}}}{{\partial k}}\right|_{k=k_i}$$ is the DOS for the $k_i$ branch of the spectrum. 

Linearizing this equation in an electric field $\bm{E}$, the distribution function can be presented in the form $f(\bm{k})=f_0(\varepsilon(k))+ f_1(\bm{k})$ with $f_1(\bm{k})= -e\bm{E}\bm{\Phi}(\bm{k})\frac{\partial f_0}{\partial\varepsilon}$. Finally we get the following equation for the vector function $\bm{\Phi}$:
\begin{equation}\label{Boltz2}
%\begin{split}
    \bm{v}_k=\sum_{i=1}^2\widetilde{D}_i(\varepsilon)N_i\int d\phi_{k_i}\left\{|\langle \bm{k}_i|V| \bm{k}\rangle |^2 + \widetilde{W}(k,k_i,\phi_{k_i}-\phi_k)\sin(\phi_{k_i}-\phi_k)\right\}[\bm{\Phi}(\bm{k}_i)-\bm{\Phi}(\bm{k})]\,,
%\end{split}	
\end{equation}
where $\widetilde{D}_i(\varepsilon)=D_i(\varepsilon)/{2\pi\hbar^2}$.
%\end{widetext}
For an axially symmetric system with the anisotropy induced by a weak electric field, the vector function $\bm{\Phi}(\bm{k})$ can be taken in the form $\bm{E}\bm{\Phi}(\bm{k})=E[\chi(k)\cos\phi_k+\varsigma(k)\sin\phi_k]$. Thus, the functions $\chi(k)$ and $\varsigma(k)$ describe amplitudes of even and odd components of the distribution function with respect to the scattering angle. To find the functions $\chi(k)$ and $\varsigma(k)$, the following equation is obtained:
\begin{equation}\label{Boltz3}
 v_k\cos\phi\!=\!\sum_{i=1}^2\widetilde{D}_i(\varepsilon)N_i\!\!\int\! d\phi_i\left\{|\langle \bm{k}_i|V| \bm{k}\rangle |^2 + \widetilde{W}(k,k_i,\phi_{k_i}-\phi_k)\sin(\phi_{k_i}-\phi_k)\right\}[\chi(k_i)\cos\phi_i+\varsigma(k_i)\sin\phi_i-\chi(k)\cos\phi_k-\varsigma(k)\sin\phi_k].
\end{equation}
%\end{widetext}
Successively multiplying equation Eq.(\ref{Boltz3}) by $\cos\phi_k$ and $\sin\phi_k$ and integrating it in each case over the interval $-\pi<\phi_k<\pi$, we obtain a system of equations for the functions $\chi(k)$ and $\varsigma(k)$:
\begin{subequations}\label{Boltz4}
   \begin{eqnarray}
     v_k&\!\!=\!\!&\sum\limits_{i=1}^2\!\widetilde{D}_i(\varepsilon)[U_1(k,k_i)\chi(k_i)-U_2(k,k_i)\chi(k)-\Delta(k,k_i)\varsigma(k_i)],\\ 
       0&\!\!=\!\!&\sum\limits_{i=1}^2\!\widetilde{D}_i(\varepsilon)[U_1(k,k_i)\varsigma(k_i)-U_2(k,k_i)\varsigma(k)+ \Delta(k,k_i)\chi(k_i)],
   \end{eqnarray}   
\end{subequations}    
\end{widetext}
where
\begin{subequations}\label{Boltz6}
\begin{eqnarray}
  U_{1}(k,k_i) = &{N_i\int d\phi|\langle \bm{k}|V| \bm{k}_i\rangle |^2\cos\phi,}\\
	U_{2}(k,k_i) = & {N_i\int d\phi|\langle \bm{k}|V| \bm{k}_i\rangle |^2,}\\
	\Delta(k,k_i)= & {N_i\int d\phi \tilde{W}(k,k_i,\phi)\sin^2\phi.}
\end{eqnarray}   
\end{subequations}

In essence, the Eqs.~(\ref{Boltz4}a,b) are a system of four equations for four variables $\chi_i(\varepsilon)=\chi(k_i(\varepsilon))$ and $\varsigma_i(\varepsilon)=\varsigma(k_i(\varepsilon))$,where $i =1,2$. It can be written as follows:  
\begin{equation}\label{set_Boltz}
\begin{pmatrix}
Q_{11}(\varepsilon)&Q_{12}(\varepsilon)&Q_{13}(\varepsilon)&Q_{14}(\varepsilon)\\
Q_{21}(\varepsilon)&Q_{22}(\varepsilon)&Q_{23}(\varepsilon)&Q_{24}(\varepsilon)\\
-Q_{13}(\varepsilon)&-Q_{14}(\varepsilon)&Q_{11}(\varepsilon)&Q_{12}(\varepsilon)\\
-Q_{23}(\varepsilon)&-Q_{24}(\varepsilon)&Q_{21}(\varepsilon)&Q_{22}(\varepsilon)\\
    \end{pmatrix}
		 \begin{pmatrix}
        \chi_1(\varepsilon)\\
       \chi_2(\varepsilon)\\
			\varsigma_1(\varepsilon)\\
			\varsigma_2(\varepsilon)
    \end{pmatrix}=
		\begin{pmatrix}
       v_{k_1}\\
        v_{k_2}\\
				0\\
				0\\
    \end{pmatrix}.
\end{equation}

The coefficients $Q_{ij}(\varepsilon)$ are given by: 
\begin{eqnarray}\label{Boltz9}
%\begin{matrix}
Q_{11}(\varepsilon)& = &\widetilde{D}_1(\varepsilon)U_{12}(k_1)-\widetilde{D}_2(\varepsilon)U_{1}(k_1,k_2)\\
Q_{22}(\varepsilon)& = &\widetilde{D}_2(\varepsilon)U_{12}(k_2)-\widetilde{D}_1(\varepsilon)U_{1}(k_2,k_1)\\
Q_{12}(\varepsilon)& = &\,\,\widetilde{D}_2(\varepsilon)U_{2}(k_1,k_2)\\
Q_{13}(\varepsilon)& = &-\widetilde{D}_1(\varepsilon)\Delta(k_1,k_1)\\
Q_{14}(\varepsilon)& = &-\widetilde{D}_2(\varepsilon)\Delta(k_1,k_2)\\
Q_{21}(\varepsilon)& = &\,\,\widetilde{D}_1(\varepsilon)U_{1}(k_1,k_1)\\
Q_{23}(\varepsilon)& = &-\widetilde{D}_1(\varepsilon)\Delta(k_2,k_1)\\
Q_{24}(\varepsilon)& = &-\widetilde{D}_2(\varepsilon)\Delta(k_2,k_2).
%\end{matrix}
\end{eqnarray}

Here we denoted $U_{12}(k) = U_{1}(k,k)-U_{2}(k,k)$.

We have solved the set of Eqs.~(\ref{set_Boltz}) numerically and found four functions $\chi_1(\varepsilon),\chi_2(\varepsilon),\varsigma_1(\varepsilon),\varsigma_2(\varepsilon)$. These functions can be considered as the components of the two two-component vectors, ($\chi_1(\varepsilon),\varsigma_1(\varepsilon))$ and  ($\chi_2(\varepsilon),\varsigma_2(\varepsilon))$, which define the two non-equilibrium distribution functions - one to each of the two Fermi contours of MHD. As all $Q_{ij} \propto N_i$ in Eq.(\ref{set_Boltz}) it is evident that all functions $\chi_1(\varepsilon),\chi_2(\varepsilon),\varsigma_1(\varepsilon),\varsigma_2(\varepsilon)$ are inversely proportional to the concentration of the scattering centers $N_i$.

Thus, for the electric field along the $OX$ axis the additions to the distribution function are: $f_{11}(\varepsilon) = -eE\chi_1(\varepsilon)\frac{\partial f_0}{\partial\varepsilon}\cos\phi$ and $f_{12}(\varepsilon)= -eE\varsigma_1(\varepsilon)\frac{\partial f_0}{\partial\varepsilon}\sin\phi$ near the inner and $f_{21}(\varepsilon)= -eE\chi_2(\varepsilon)\frac{\partial f_0}{\partial\varepsilon}\cos\phi$ and $f_{22}(\varepsilon)= -eE\varsigma_2(\varepsilon)\frac{\partial f_0}{\partial\varepsilon}\sin\phi$ near the outer Fermi contours. 

Spin-up and spin-down electrons create equal currents along the electric field, which can be written as:
\begin{equation}\label{current_x}
\begin{split}
J_x=\frac{e^2E}{\hbar^2}\int\frac{d\phi d\varepsilon}{(2\pi)^2} \cos^2{\phi}\frac{\partial f_0}{\partial\varepsilon}\times \\ \left [D_1(\varepsilon)\chi_1(\varepsilon)\frac{\partial{\varepsilon_{k}}}{{\partial k}}|_{k=k_1}+D_2(\varepsilon)\chi_2(\varepsilon)\frac{\partial{\varepsilon_{k}}}{{\partial k}}|_{k=k_2}\right].
\end{split}
\end{equation}
The total current is practically the same as what was calculated previously~\cite{Shchamkhalova}, since the corrections due to higher order  terms in $V$   are small.

The electron current in the transverse to the electric field direction has two components. One of them, caused by the skew scattering processes, is determined by the velocity component $\partial \varepsilon /\partial k_y$ and by asymmetric in scattering angle additions to the distribution function, $f_{1 2}(\varepsilon)$ and $f_{22}(\varepsilon)$.  For spin-up electrons this extrinsic SHC current equals:
\begin{equation}
\begin{split}
J_y= \frac{e^2E}{\hbar^2}\int\frac{d\phi d\varepsilon}{(2\pi)^2} \sin^2{\phi}\frac{\partial f_0}{\partial\varepsilon}\times \\ \left[D_1(\varepsilon)\varsigma_1(\varepsilon)\frac{\partial{\varepsilon_{k}}}{{\partial k}}|_{k=k_1}+D_2(\varepsilon)\varsigma_2(\varepsilon)\frac{\partial{\varepsilon_{k}}}{{\partial k}}|_{k=k_2}\right].
\end{split}
\end{equation}
Spin-down electrons produce the current of opposed sign, so that the total electric current in $y$ direction is absent. However, the spin currents add up, and the total extrinsic SHC equals $$J_{ext}^{s}=-2(\hbar/e) J_y.$$
\begin{figure}  
		\centerline{\includegraphics[width=1\linewidth]{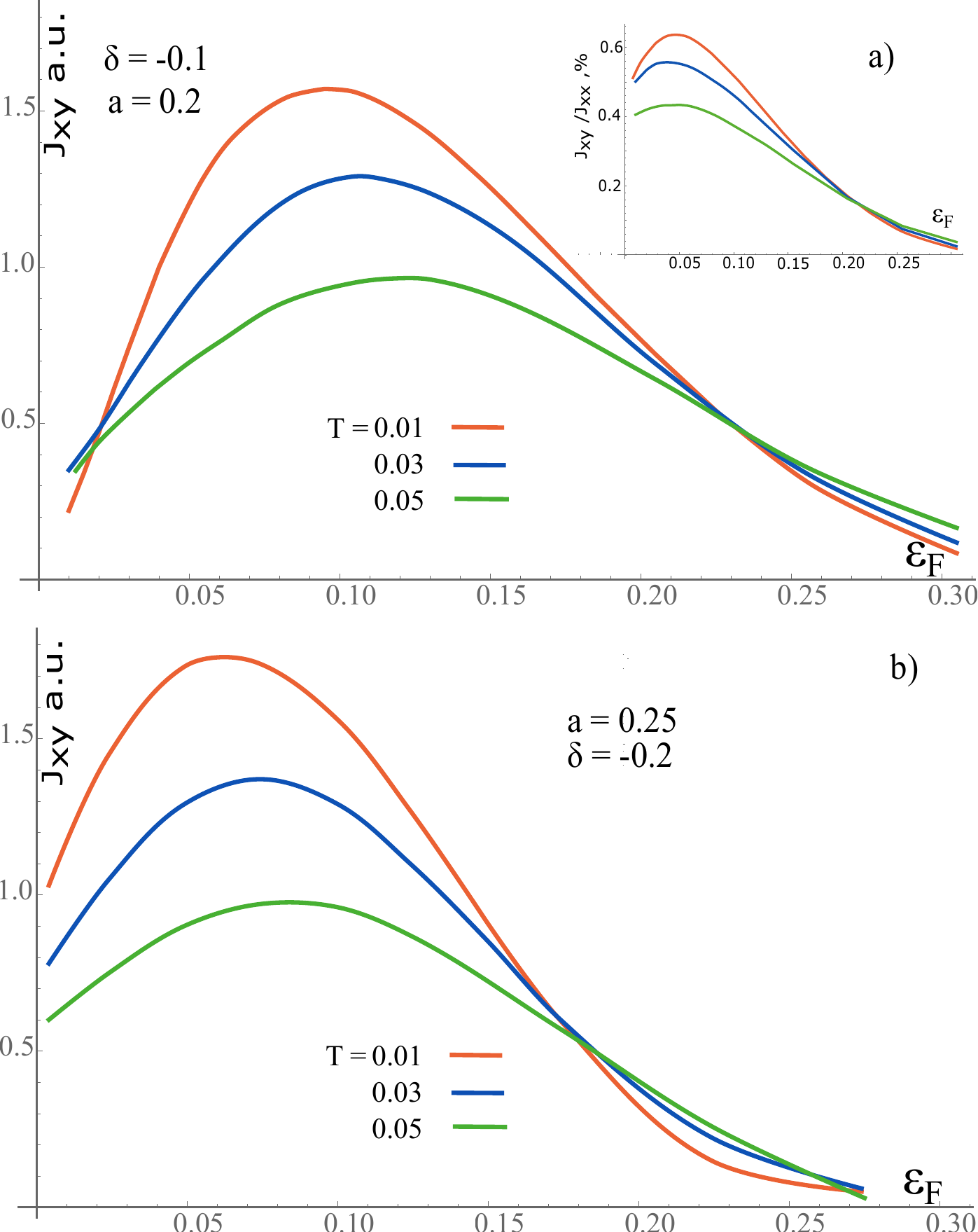}}
    \caption{Extrinsic SHC versus $E_F$ for two systems: 1) $a = 0.2, \delta=-0.1$ (a); 2) $a = 0.25, \delta = -0.2$ (b) and three temperatures, $N_i=1$. Inset shows the Hall angle versus $E_F$ for the first system. }
\label{f2:fig2}	
\end{figure} 

\begin{figure}   
		\centerline{\includegraphics[width=1\linewidth]{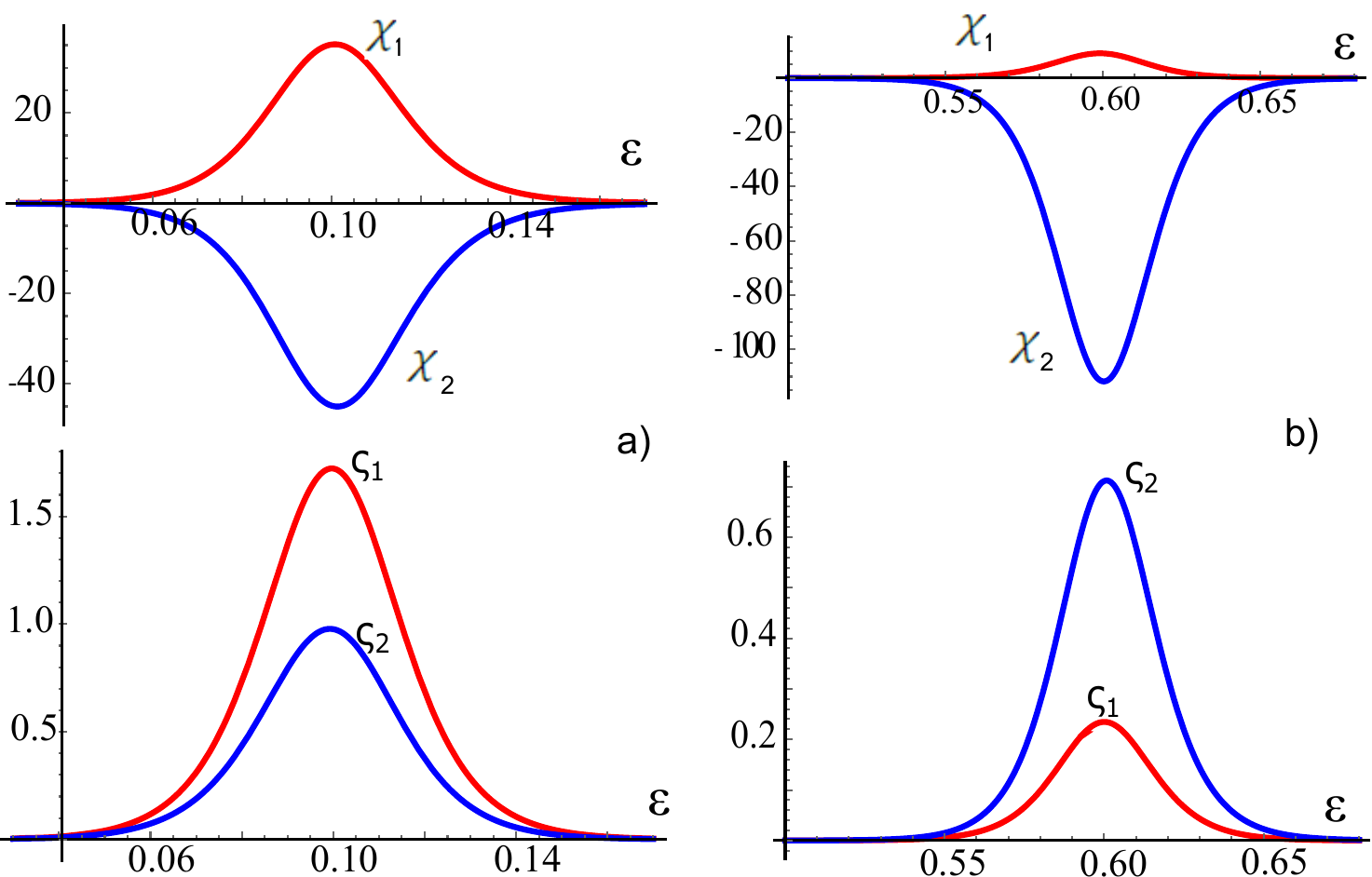}}
    \caption{The additions to the distribution function for two $E_F$:  a) $E_F= 0.1$ above the conduction band minimum; b)$E_F= 0.6$ (about 0.1 lower than the Mexican hat top). The symmetric (upper curves) and asymmetric (lower curves) in scattering angle parts are shown  at temperature T = 0.01, $N_i=1$ .}
 \label{f3:fig3}
\end{figure} 
	
\begin{figure}[htbp]
    \centerline{\includegraphics[width=0.8\linewidth]{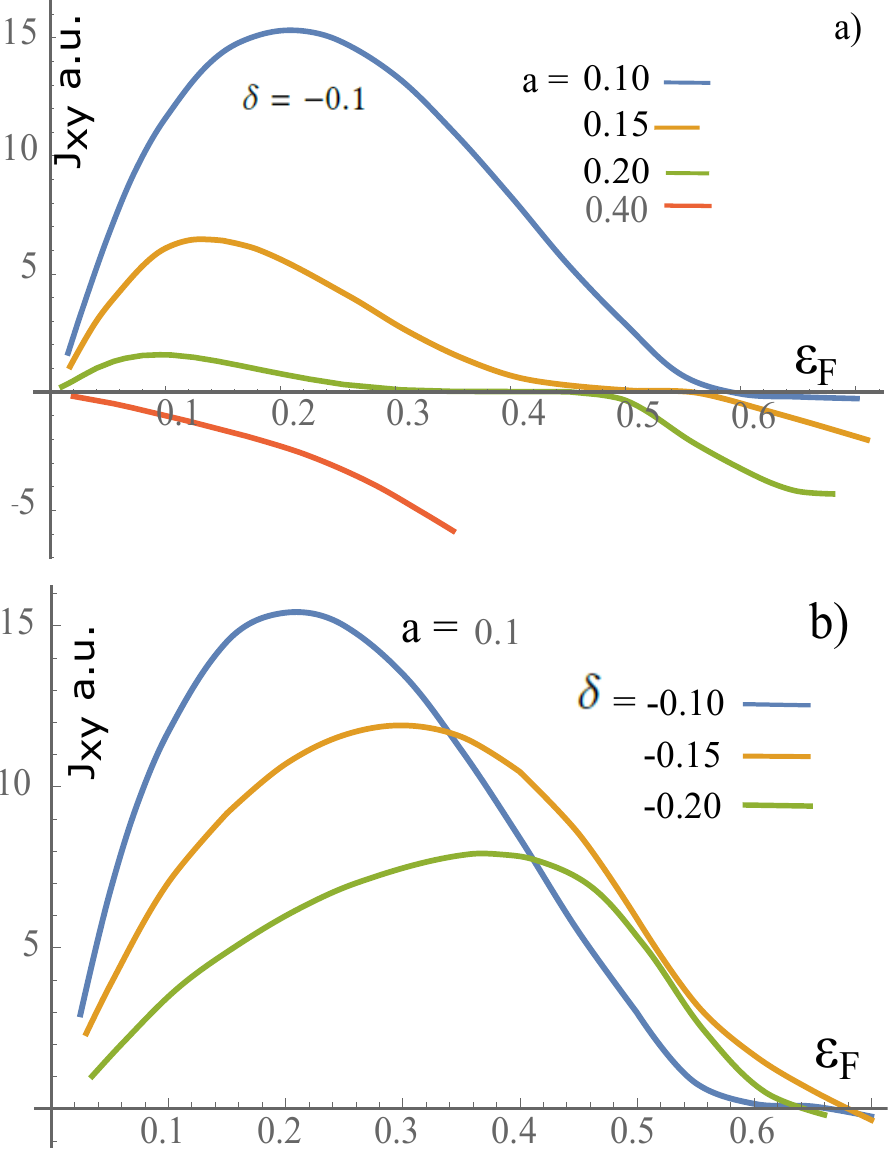}}
    \caption{Extrinsic SHC dependencies on $E_F$ for: a) $\delta = -0.1$ and different values of parameter $a$; b) $a=0.1$ and  $\delta = -0.1; -0.15; -0.2$ at temperature T = 0.01, $N_i = 1$}
    \label{f4:fig4}
\end{figure}

Fig.~\ref{f2:fig2} presents the dependencies of the extrinsic SHC for two systems with parameters $ a = 0.2, \delta = -0.1$ and $ a = 0.25, \delta = -0.2$ on the $E_F$ for different temperatures. These dependencies are similar for both systems. At the low $E_F$ the extrinsic SHC increases with the $E_F$. We attribute this to an increase of the carrier number. But with further increase of the $E_F$ the SHC reaches maximum and then decreases and even changes the sign. The temperature dependencies of the SHC and the current along electric field are similar: both decrease with temperature at low $E_F$. This is a result of the decrease of the screening of charge impurity by electrons with increasing temperature~\cite{Shchamkhalova}.  

The inset in Fig.~\ref{f2:fig2} shows that at the low $E_F$ the extrinsic SHC current is approximately two orders of magnitude smaller than the longitudinal charge current, although calculations show that $\varsigma_i$ is only one order of magnitude smaller than $\chi_i$ (see Fig.~\ref{f3:fig3}). This is explained by the fact that both $\chi_1$, $\chi_2$ and the electron velocities near the two isoenergetic contours have opposite signs, so the charge currents near the two contours are summed. But $\varsigma_1$ and $\varsigma_2$ have the same signs near both contours when the electron velocities near the two contours have opposite signs. As a result, the extrinsic SHC near these contours have opposite signs and are subtracted. Thus, the resulting extrinsic SHC is significantly smaller than the longitudinal charge current. 

Moreover, at the low $E_F$ the skew scattering probability near the inner contour is larger than near the outer contour, hence $\varsigma_1 > \varsigma_2$, although $\chi_1 < \chi_2$ (see Fig.~\ref{f3:fig3}a)). So the SHC due to electrons near the inner contour is prevailing at low $E_F$ and these electrons determined the direction of the SHC. With increasing $E_F$, as it approaches the top of the mexican hat,  the velocity of electrons near the inner contour decreases rapidly. As a result both $\chi_1$ and $\varsigma_1$ are also much smaller than $\chi_2$ and $\varsigma_2$ (Fig.~\ref{f3:fig3}b)). Now the contribution  of electrons near the outer Fermi contour into the extrinsic SHC becomes predominant and the SHC changes the sign - it is directed along the spin current of the electrons near the outer contour. Thus the extrinsic SHC changes the sign at some value of the Fermi level, which depends on the profile of the MHD. The latter is known to determine by two parameters of the model: one is the parameter $a$ describing the hybridization of the electrons and hole bands  and the other is $\delta$  defining the asymmetry of these bands.

Fig.~\ref{f4:fig4}a) shows the dependencies of the extrinsic SHC on the $E_F$ for systems with different hybridization parameters $a$ and the same parameter $\delta = -0.1$ and temperature $T = 0.01$. It is clearly seen that the value of the extrinsic SHC decreases with the increase of $a$. For $\delta= -0.1$ and $a < 0.25$ the SHC changes the sign at $E_F > 0.5$. At $ a > 0.3$ the SHC is negative for all values of $E_F < 1$.

The dependencies of the extrinsic SHC on the $E_F$ for the systems with the hybridization parameters $a = 0.1$ and various $\delta$: $\delta = -0.1; -0.15; -0.2$ at a temperature $T = 0.01$ are shown in fig.\ref{f4:fig4}b).  The extrinsic SHC decreases with decreasing of $\delta$ at fixed $a = 0.1$.

\begin{figure}
    \centerline{\includegraphics[width=1\linewidth]{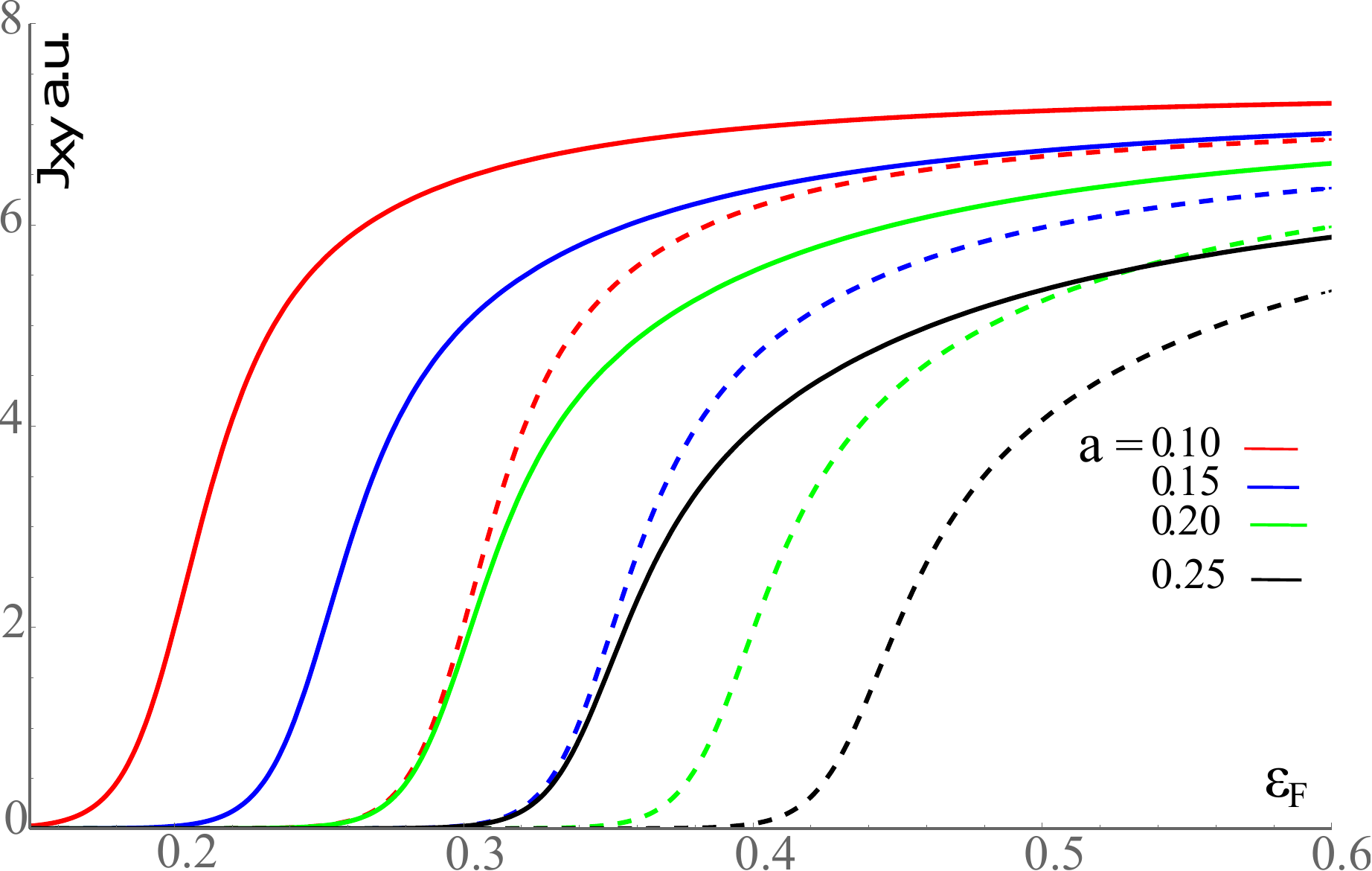}}
    \caption{Intrinsic SHC dependencies on $E_F$ for three values of parameter $a$ and two values of $\delta$: $\delta$ = -0.1 (thick curves) and $\delta$ = -0.2 (dashed curves). Dashed curves are shifted for clarity.}
\label{f5:fig5}		
 \end{figure}

The intrinsic component of the SHC is defined by the velocity component caused by the Berry curvature. Fig.~\ref{f5:fig5}  shows the dependencies of the intrinsic SHC on the $E_F$ for different values of the hybridization parameter $a$ and two values of band asymmetry $\delta$: $\delta = - 0.1$ (thick curves) and $\delta = - 0.2$ (dashed curves), for temperature, $T = 0.01$ and $N_i=1$. With increasing $E_F$ the intrinsic SHC increases rapidly and then saturates.  It is seen that the intrinsic SHC decreases with the increase of $a$ as does the extrinsic SHC. The dependence of intrinsic SHC on $\delta$ is the same for all values of $a$ - the intrinsic SHC decreases with the increase of a band asymmetry. Although the Berry curvature (Eq.(\ref{Berry}))  is independent of $\delta$, the intrinsic SHC depends on $\delta$ as does the DOS. 

\begin{figure}
    \centerline{\includegraphics[width=1\linewidth]{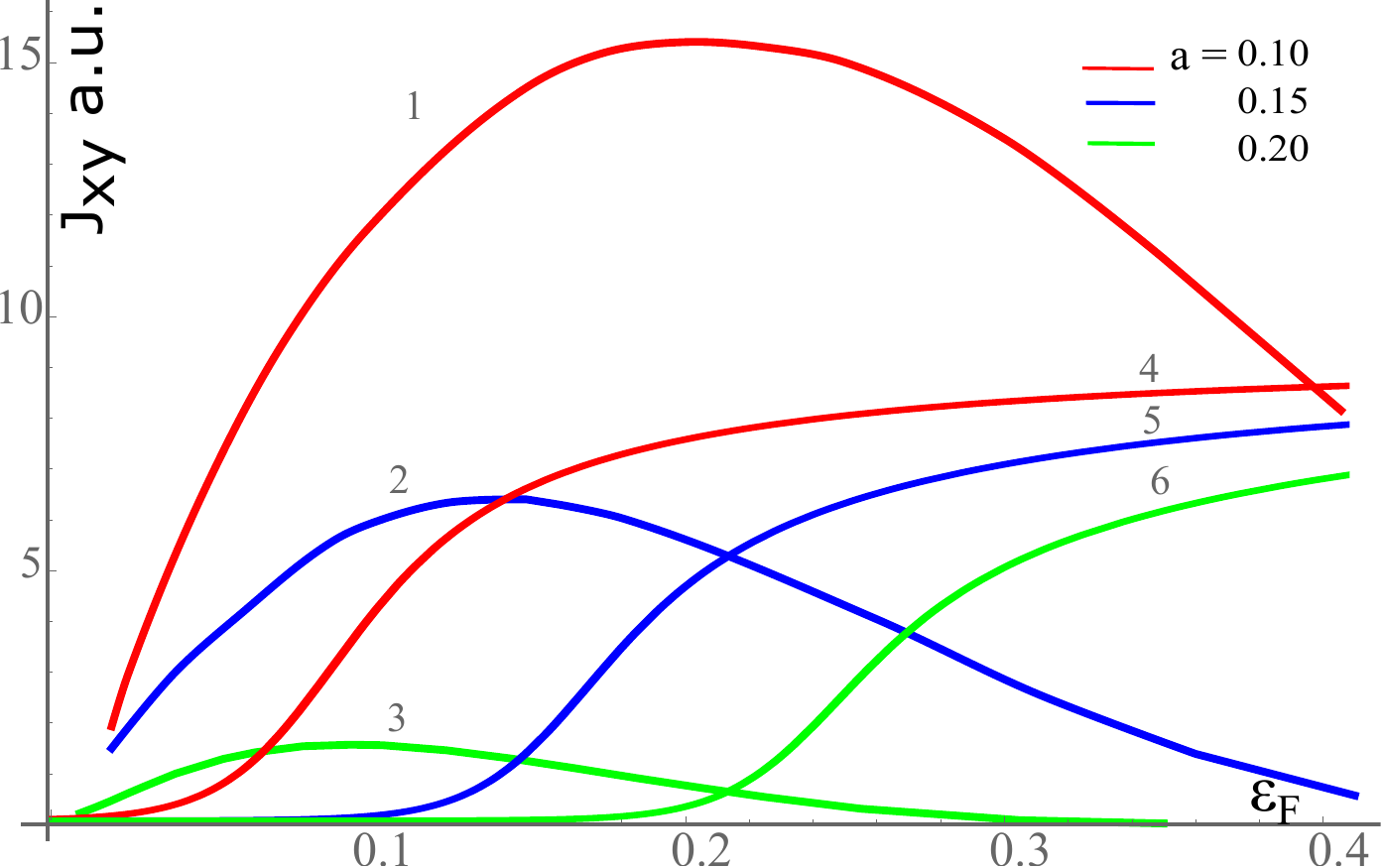}}
    \caption{Extrinsic (curves 1-3) and intrinsic (curves 4-6) SHC dependencies on $E_F$ for $\delta$ = -0.1 and three values of parameter $a$. Curves 5 and 6 are shifted for clarity.}
 \label{f6:fig6}
\end{figure}

Fig.~\ref{f6:fig6}  shows the dependencies of the intrinsic and extrinsic SHC for $\delta$ = -0.1 and three values of parameter $a$  on the $E_F$ at a temperature $T = 0.01$ and $N_i = 1$. The extrinsic SHC due to skew scattering by charge impurities is of the order of or larger than the intrinsic one at low hybridization parameter $a$ and low $E_F$. The extrinsic SHC will be more larger at smaller $N_i$, $N_i < 1$, because the extrinsic SHC increases with the decrease of the impurity concentration $N_i$. But for very small $N_i$ our approach based on a small perturbation of the Fermi distribution function will be not applicable.

\section{Conclusion}\label{S_conclusion}
In conclusion, we investigated the spin Hall effect in topological insulators with MHD, which arises due to the inversion of the electron and hole bands. The SHC has two components: the extrinsic SHC, caused by the spin-dependent scattering of electrons by charged impurities, and the intrinsic SHC, arising from the Berry curvature. The presence of MHD significantly affects both components .

The extrinsic SHC changes as a result of a radical rearrangement of scattering processes and the important role of the DOS singularity at the MHD bottom. The main effect is due to the doubly connected geometry of the Fermi surface and the corresponding appearance of two Fermi contours, the electron transitions between which create an effective scattering channel absent in conventional systems. This scattering channel leads to a radical rearrangement of the screening charge and a significant change in the scattering probability, mobility, and asymmetric spin-dependent scattering. These effects are greatest when the $E_F$ lies near the MHD bottom.

We found that at low $E_F$ the main contribution into extrinsic SHC is that of electrons near the inner Fermi contour. With the increase of the $E_F$ the extrinsic SHC first increases due to the increase of a carrier number and the extrinsic SHC can significantly exceed the intrinsic SHC. Then the decrease of the skew scattering probability with increasing electron energy and the decrease of the velocity of the electrons near the inner Fermi contour, when the Fermi level is close to the Mexican hat top, become the prevailing factors, which lead to the decrease of the SHC. With further increase of the $E_F$ the main contribution into SHC is that of electrons near the outer Fermi contour, which velocity sign is opposed to that of electrons near the inner contour, and SHC changes the sign. We believe that such change of the SHC sign is a special property of the systems with a spectrum of MHD type.

The intrinsic SHC also increases rapidly near the MHD bottom due to the increase in the Berry curvature. We compared both components for InAs/InGaSb quantum wells with a density of charged impurities $N_i= 1$. It was found that the extrinsic SHC can dominate for the $E_F$ near the MHD bottom, but at high $E_F$, the intrinsic SHC significantly dominates. 

Both extrinsic and intrinsic components of SHC decrease with the increase of the hybridization parameter $a$, what leads to a flattening of the MHD profile, confirming the important role of MHD-type spectrum in the effects under consideration.   

\begin{acknowledgments}

This work was carried out in the framework of the state task (theme code FFWZ-2025-0014) for the Kotelnikov Institute of Radio Engineering and Electronics.
\end{acknowledgments}

%\bibliography{skew_may}

\begin{thebibliography}{33}%
\makeatletter
\providecommand \@ifxundefined [1]{%
 \@ifx{#1\undefined}
}%
\providecommand \@ifnum [1]{%
 \ifnum #1\expandafter \@firstoftwo
 \else \expandafter \@secondoftwo
 \fi
}%
\providecommand \@ifx [1]{%
 \ifx #1\expandafter \@firstoftwo
 \else \expandafter \@secondoftwo
 \fi
}%
\providecommand \natexlab [1]{#1}%
\providecommand \enquote  [1]{``#1''}%
\providecommand \bibnamefont  [1]{#1}%
\providecommand \bibfnamefont [1]{#1}%
\providecommand \citenamefont [1]{#1}%
\providecommand \href@noop [0]{\@secondoftwo}%
\providecommand \href [0]{\begingroup \@sanitize@url \@href}%
\providecommand \@href[1]{\@@startlink{#1}\@@href}%
\providecommand \@@href[1]{\endgroup#1\@@endlink}%
\providecommand \@sanitize@url [0]{\catcode `\\12\catcode `\$12\catcode
  `\&12\catcode `\#12\catcode `\^12\catcode `\_12\catcode `\%12\relax}%
\providecommand \@@startlink[1]{}%
\providecommand \@@endlink[0]{}%
\providecommand \url  [0]{\begingroup\@sanitize@url \@url }%
\providecommand \@url [1]{\endgroup\@href {#1}{\urlprefix }}%
\providecommand \urlprefix  [0]{URL }%
\providecommand \Eprint [0]{\href }%
\providecommand \doibase [0]{http://dx.doi.org/}%
\providecommand \selectlanguage [0]{\@gobble}%
\providecommand \bibinfo  [0]{\@secondoftwo}%
\providecommand \bibfield  [0]{\@secondoftwo}%
\providecommand \translation [1]{[#1]}%
\providecommand \BibitemOpen [0]{}%
\providecommand \bibitemStop [0]{}%
\providecommand \bibitemNoStop [0]{.\EOS\space}%
\providecommand \EOS [0]{\spacefactor3000\relax}%
\providecommand \BibitemShut  [1]{\csname bibitem#1\endcsname}%
\let\auto@bib@innerbib\@empty
%</preamble>
\bibitem [{\citenamefont {Sinova}\ \emph {et~al.}(2015)\citenamefont {Sinova},
  \citenamefont {Valenzuela}, \citenamefont {Wunderlich}, \citenamefont
  {Back},\ and\ \citenamefont {Jungwirth}}]{Sinova}%
  \BibitemOpen
  \bibfield  {author} {\bibinfo {author} {\bibfnamefont {J.}~\bibnamefont
  {Sinova}}, \bibinfo {author} {\bibfnamefont {S.~O.}\ \bibnamefont
  {Valenzuela}}, \bibinfo {author} {\bibfnamefont {J.}~\bibnamefont
  {Wunderlich}}, \bibinfo {author} {\bibfnamefont {C.~H.}\ \bibnamefont
  {Back}}, \ and\ \bibinfo {author} {\bibfnamefont {T.}~\bibnamefont
  {Jungwirth}},\ }\href {\doibase 10.1103/RevModPhys.87.1213} {\bibfield
  {journal} {\bibinfo  {journal} {Rev. Mod. Phys.}\ }\textbf {\bibinfo {volume}
  {87}},\ \bibinfo {pages} {1213} (\bibinfo {year} {2015})}\BibitemShut
  {NoStop}%
\bibitem [{\citenamefont {Sinitsyn}(2007)}]{Sinitsyn_2008}%
  \BibitemOpen
  \bibfield  {author} {\bibinfo {author} {\bibfnamefont {N.~A.}\ \bibnamefont
  {Sinitsyn}},\ }\href {\doibase 10.1088/0953-8984/20/02/023201} {\bibfield
  {journal} {\bibinfo  {journal} {Journal of Physics: Condensed Matter}\
  }\textbf {\bibinfo {volume} {20}},\ \bibinfo {pages} {023201} (\bibinfo
  {year} {2007})}\BibitemShut {NoStop}%
\bibitem [{\citenamefont {Chang}\ \emph {et~al.}(2023)\citenamefont {Chang},
  \citenamefont {Liu},\ and\ \citenamefont {MacDonald}}]{RevModPhys.95.011002}%
  \BibitemOpen
  \bibfield  {author} {\bibinfo {author} {\bibfnamefont {C.-Z.}\ \bibnamefont
  {Chang}}, \bibinfo {author} {\bibfnamefont {C.-X.}\ \bibnamefont {Liu}}, \
  and\ \bibinfo {author} {\bibfnamefont {A.~H.}\ \bibnamefont {MacDonald}},\
  }\href {\doibase 10.1103/RevModPhys.95.011002} {\bibfield  {journal}
  {\bibinfo  {journal} {Rev. Mod. Phys.}\ }\textbf {\bibinfo {volume} {95}},\
  \bibinfo {pages} {011002} (\bibinfo {year} {2023})}\BibitemShut {NoStop}%
\bibitem [{\citenamefont {Niu}\ \emph {et~al.}(2020)\citenamefont {Niu},
  \citenamefont {Wang}, \citenamefont {Mao}, \citenamefont {Huang},
  \citenamefont {Mokrousov},\ and\ \citenamefont
  {Dai}}]{PhysRevLett.124.066401}%
  \BibitemOpen
  \bibfield  {author} {\bibinfo {author} {\bibfnamefont {C.}~\bibnamefont
  {Niu}}, \bibinfo {author} {\bibfnamefont {H.}~\bibnamefont {Wang}}, \bibinfo
  {author} {\bibfnamefont {N.}~\bibnamefont {Mao}}, \bibinfo {author}
  {\bibfnamefont {B.}~\bibnamefont {Huang}}, \bibinfo {author} {\bibfnamefont
  {Y.}~\bibnamefont {Mokrousov}}, \ and\ \bibinfo {author} {\bibfnamefont
  {Y.}~\bibnamefont {Dai}},\ }\href {\doibase 10.1103/PhysRevLett.124.066401}
  {\bibfield  {journal} {\bibinfo  {journal} {Phys. Rev. Lett.}\ }\textbf
  {\bibinfo {volume} {124}},\ \bibinfo {pages} {066401} (\bibinfo {year}
  {2020})}\BibitemShut {NoStop}%
\bibitem [{\citenamefont {Wang}\ \emph {et~al.}(2025)\citenamefont {Wang},
  \citenamefont {Liu}, \citenamefont {Feng}, \citenamefont {Cao}, \citenamefont
  {Wu}, \citenamefont {Lai}, \citenamefont {Gao}, \citenamefont {Xiao},\ and\
  \citenamefont {Yang}}]{PhysRevLett.134.056301}%
  \BibitemOpen
  \bibfield  {author} {\bibinfo {author} {\bibfnamefont {H.}~\bibnamefont
  {Wang}}, \bibinfo {author} {\bibfnamefont {H.}~\bibnamefont {Liu}}, \bibinfo
  {author} {\bibfnamefont {X.}~\bibnamefont {Feng}}, \bibinfo {author}
  {\bibfnamefont {J.}~\bibnamefont {Cao}}, \bibinfo {author} {\bibfnamefont
  {W.}~\bibnamefont {Wu}}, \bibinfo {author} {\bibfnamefont {S.}~\bibnamefont
  {Lai}}, \bibinfo {author} {\bibfnamefont {W.}~\bibnamefont {Gao}}, \bibinfo
  {author} {\bibfnamefont {C.}~\bibnamefont {Xiao}}, \ and\ \bibinfo {author}
  {\bibfnamefont {S.~A.}\ \bibnamefont {Yang}},\ }\href {\doibase
  10.1103/PhysRevLett.134.056301} {\bibfield  {journal} {\bibinfo  {journal}
  {Phys. Rev. Lett.}\ }\textbf {\bibinfo {volume} {134}},\ \bibinfo {pages}
  {056301} (\bibinfo {year} {2025})}\BibitemShut {NoStop}%
\bibitem [{\citenamefont {Hayami}\ \emph {et~al.}(2022)\citenamefont {Hayami},
  \citenamefont {Yatsushiro},\ and\ \citenamefont
  {Kusunose}}]{PhysRevB.106.024405}%
  \BibitemOpen
  \bibfield  {author} {\bibinfo {author} {\bibfnamefont {S.}~\bibnamefont
  {Hayami}}, \bibinfo {author} {\bibfnamefont {M.}~\bibnamefont {Yatsushiro}},
  \ and\ \bibinfo {author} {\bibfnamefont {H.}~\bibnamefont {Kusunose}},\
  }\href {\doibase 10.1103/PhysRevB.106.024405} {\bibfield  {journal} {\bibinfo
   {journal} {Phys. Rev. B}\ }\textbf {\bibinfo {volume} {106}},\ \bibinfo
  {pages} {024405} (\bibinfo {year} {2022})}\BibitemShut {NoStop}%
\bibitem [{\citenamefont {Manchon}\ \emph {et~al.}(2019)\citenamefont
  {Manchon}, \citenamefont {\ifmmode~\check{Z}\else \v{Z}\fi{}elezn\'y},
  \citenamefont {Miron}, \citenamefont {Jungwirth}, \citenamefont {Sinova},
  \citenamefont {Thiaville}, \citenamefont {Garello},\ and\ \citenamefont
  {Gambardella}}]{Manchon1}%
  \BibitemOpen
  \bibfield  {author} {\bibinfo {author} {\bibfnamefont {A.}~\bibnamefont
  {Manchon}}, \bibinfo {author} {\bibfnamefont {J.}~\bibnamefont
  {\ifmmode~\check{Z}\else \v{Z}\fi{}elezn\'y}}, \bibinfo {author}
  {\bibfnamefont {I.~M.}\ \bibnamefont {Miron}}, \bibinfo {author}
  {\bibfnamefont {T.}~\bibnamefont {Jungwirth}}, \bibinfo {author}
  {\bibfnamefont {J.}~\bibnamefont {Sinova}}, \bibinfo {author} {\bibfnamefont
  {A.}~\bibnamefont {Thiaville}}, \bibinfo {author} {\bibfnamefont
  {K.}~\bibnamefont {Garello}}, \ and\ \bibinfo {author} {\bibfnamefont
  {P.}~\bibnamefont {Gambardella}},\ }\href {\doibase
  10.1103/RevModPhys.91.035004} {\bibfield  {journal} {\bibinfo  {journal}
  {Rev. Mod. Phys.}\ }\textbf {\bibinfo {volume} {91}},\ \bibinfo {pages}
  {035004} (\bibinfo {year} {2019})}\BibitemShut {NoStop}%
\bibitem [{\citenamefont {Bai}\ \emph {et~al.}(2026)\citenamefont {Bai},
  \citenamefont {Yuan}, \citenamefont {Chen}, \citenamefont {Dai},
  \citenamefont {Huang}, \citenamefont {Wang},\ and\ \citenamefont
  {Niu}}]{588c-z8cy}%
  \BibitemOpen
  \bibfield  {author} {\bibinfo {author} {\bibfnamefont {Y.}~\bibnamefont
  {Bai}}, \bibinfo {author} {\bibfnamefont {B.}~\bibnamefont {Yuan}}, \bibinfo
  {author} {\bibfnamefont {Z.}~\bibnamefont {Chen}}, \bibinfo {author}
  {\bibfnamefont {Y.}~\bibnamefont {Dai}}, \bibinfo {author} {\bibfnamefont
  {B.}~\bibnamefont {Huang}}, \bibinfo {author} {\bibfnamefont
  {X.}~\bibnamefont {Wang}}, \ and\ \bibinfo {author} {\bibfnamefont
  {C.}~\bibnamefont {Niu}},\ }\href {\doibase 10.1103/588c-z8cy} {\bibfield
  {journal} {\bibinfo  {journal} {Phys. Rev. Lett.}\ }\textbf {\bibinfo
  {volume} {136}},\ \bibinfo {pages} {046602} (\bibinfo {year}
  {2026})}\BibitemShut {NoStop}%
\bibitem [{\citenamefont {Bernevig}\ and\ \citenamefont
  {Zhang}(2006)}]{PhysRevLett.96.106802}%
  \BibitemOpen
  \bibfield  {author} {\bibinfo {author} {\bibfnamefont {B.~A.}\ \bibnamefont
  {Bernevig}}\ and\ \bibinfo {author} {\bibfnamefont {S.-C.}\ \bibnamefont
  {Zhang}},\ }\href {\doibase 10.1103/PhysRevLett.96.106802} {\bibfield
  {journal} {\bibinfo  {journal} {Phys. Rev. Lett.}\ }\textbf {\bibinfo
  {volume} {96}},\ \bibinfo {pages} {106802} (\bibinfo {year}
  {2006})}\BibitemShut {NoStop}%
\bibitem [{\citenamefont {Jiabin}\ \emph {et~al.}(2025)\citenamefont {Jiabin},
  \citenamefont {Bernevig B.~Andrei}, \citenamefont {Rossi~Enrico},\ and\
  \citenamefont {Bohm-Jung}}]{Yu2025}%
  \BibitemOpen
  \bibfield  {author} {\bibinfo {author} {\bibfnamefont {Y.}~\bibnamefont
  {Jiabin}}, \bibinfo {author} {\bibfnamefont {Q.~R.}\ \bibnamefont {Bernevig
  B.~Andrei}}, \bibinfo {author} {\bibfnamefont {T.~P.}\ \bibnamefont
  {Rossi~Enrico}}, \ and\ \bibinfo {author} {\bibfnamefont {Y.}~\bibnamefont
  {Bohm-Jung}},\ }\href {\doibase 10.1038/s41535-025-00801-3} {\bibfield
  {journal} {\bibinfo  {journal} {Quantum Materials}\ }\textbf {\bibinfo
  {volume} {10}},\ \bibinfo {pages} {1} (\bibinfo {year} {2025})}\BibitemShut
  {NoStop}%
\bibitem [{\citenamefont {Hirsch}(1999)}]{Hirsch}%
  \BibitemOpen
  \bibfield  {author} {\bibinfo {author} {\bibfnamefont {J.~E.}\ \bibnamefont
  {Hirsch}},\ }\href {\doibase 10.1103/PhysRevLett.83.1834} {\bibfield
  {journal} {\bibinfo  {journal} {Phys. Rev. Lett.}\ }\textbf {\bibinfo
  {volume} {83}},\ \bibinfo {pages} {1834} (\bibinfo {year}
  {1999})}\BibitemShut {NoStop}%
\bibitem [{\citenamefont {R}\ \emph {et~al.}(2014)\citenamefont {R},
  \citenamefont {Lee J.~S.}, \citenamefont {Fischer M.~H.},\ and\ \citenamefont
  {Manchon~A.}}]{Manchon}%
  \BibitemOpen
  \bibfield  {author} {\bibinfo {author} {\bibfnamefont {M.~A.}\ \bibnamefont
  {R}}, \bibinfo {author} {\bibfnamefont {M.~P.~J.}\ \bibnamefont {Lee J.~S.},
  \bibfnamefont {Richardella A.and Grab J.~L.}}, \bibinfo {author}
  {\bibfnamefont {V.~A.}\ \bibnamefont {Fischer M.~H.}}, \ and\ \bibinfo
  {author} {\bibfnamefont {R.~D.~C.}\ \bibnamefont {Manchon~A.}, \bibfnamefont
  {Kim E.-A.and Samarth~N.}},\ }\href {\doibase 10.1038/nature13534} {\bibfield
   {journal} {\bibinfo  {journal} {Nature}\ }\textbf {\bibinfo {volume}
  {511}},\ \bibinfo {pages} {449} (\bibinfo {year} {2014})}\BibitemShut
  {NoStop}%
\bibitem [{\citenamefont {Krishtopenko}(2021)}]{Krishtopenko}%
  \BibitemOpen
  \bibfield  {author} {\bibinfo {author} {\bibfnamefont {S.~S.}\ \bibnamefont
  {Krishtopenko}},\ }\href {\doibase 10.1038/s41598-021-00577-z} {\bibfield
  {journal} {\bibinfo  {journal} {Scientific Reports}\ }\textbf {\bibinfo
  {volume} {11}},\ \bibinfo {pages} {21060} (\bibinfo {year}
  {2021})}\BibitemShut {NoStop}%
\bibitem [{\citenamefont {Stauber}\ \emph {et~al.}(2007)\citenamefont
  {Stauber}, \citenamefont {Peres}, \citenamefont {Guinea},\ and\ \citenamefont
  {Castro~Neto}}]{Stauber}%
  \BibitemOpen
  \bibfield  {author} {\bibinfo {author} {\bibfnamefont {T.}~\bibnamefont
  {Stauber}}, \bibinfo {author} {\bibfnamefont {N.~M.~R.}\ \bibnamefont
  {Peres}}, \bibinfo {author} {\bibfnamefont {F.}~\bibnamefont {Guinea}}, \
  and\ \bibinfo {author} {\bibfnamefont {A.~H.}\ \bibnamefont {Castro~Neto}},\
  }\href {\doibase 10.1103/PhysRevB.75.115425} {\bibfield  {journal} {\bibinfo
  {journal} {Phys. Rev. B}\ }\textbf {\bibinfo {volume} {75}},\ \bibinfo
  {pages} {115425} (\bibinfo {year} {2007})}\BibitemShut {NoStop}%
\bibitem [{\citenamefont {Wickramaratne}\ \emph {et~al.}(2015)\citenamefont
  {Wickramaratne}, \citenamefont {Zahid},\ and\ \citenamefont
  {Lake}}]{10.1063/1.4928559}%
  \BibitemOpen
  \bibfield  {author} {\bibinfo {author} {\bibfnamefont {D.}~\bibnamefont
  {Wickramaratne}}, \bibinfo {author} {\bibfnamefont {F.}~\bibnamefont
  {Zahid}}, \ and\ \bibinfo {author} {\bibfnamefont {R.~K.}\ \bibnamefont
  {Lake}},\ }\href {\doibase 10.1063/1.4928559} {\bibfield  {journal} {\bibinfo
   {journal} {Journal of Applied Physics}\ }\textbf {\bibinfo {volume} {118}},\
  \bibinfo {pages} {075101} (\bibinfo {year} {2015})},\ \Eprint
  {http://arxiv.org/abs/https://pubs.aip.org/aip/jap/article-pdf/doi/10.1063/1.4928559/14743158/075101\_1\_online.pdf}
  {https://pubs.aip.org/aip/jap/article-pdf/doi/10.1063/1.4928559/14743158/075101\_1\_online.pdf}
  \BibitemShut {NoStop}%
\bibitem [{\citenamefont {Kremer}\ \emph {et~al.}(2025)\citenamefont {Kremer},
  \citenamefont {Mahmoudi}, \citenamefont {Bouaziz}, \citenamefont {Rahimi},
  \citenamefont {Bertran}, \citenamefont {Dayen}, \citenamefont {Rocca},
  \citenamefont {Pala}, \citenamefont {Naitabdi}, \citenamefont {Chaste},
  \citenamefont {Oehler},\ and\ \citenamefont {Ouerghi}}]{kremer2025}%
  \BibitemOpen
  \bibfield  {author} {\bibinfo {author} {\bibfnamefont {G.}~\bibnamefont
  {Kremer}}, \bibinfo {author} {\bibfnamefont {A.}~\bibnamefont {Mahmoudi}},
  \bibinfo {author} {\bibfnamefont {M.}~\bibnamefont {Bouaziz}}, \bibinfo
  {author} {\bibfnamefont {M.}~\bibnamefont {Rahimi}}, \bibinfo {author}
  {\bibfnamefont {F.}~\bibnamefont {Bertran}}, \bibinfo {author} {\bibfnamefont
  {J.-F.}\ \bibnamefont {Dayen}}, \bibinfo {author} {\bibfnamefont {M.~L.~D.}\
  \bibnamefont {Rocca}}, \bibinfo {author} {\bibfnamefont {M.}~\bibnamefont
  {Pala}}, \bibinfo {author} {\bibfnamefont {A.}~\bibnamefont {Naitabdi}},
  \bibinfo {author} {\bibfnamefont {J.}~\bibnamefont {Chaste}}, \bibinfo
  {author} {\bibfnamefont {F.}~\bibnamefont {Oehler}}, \ and\ \bibinfo {author}
  {\bibfnamefont {A.}~\bibnamefont {Ouerghi}},\ }\href
  {https://arxiv.org/abs/2509.06488} {\enquote {\bibinfo {title} {Mexican
  hat-like valence band dispersion and quantum confinement in rhombohedral
  ferroelectric alpha-in2se3},}\ } (\bibinfo {year} {2025}),\ \Eprint
  {http://arxiv.org/abs/2509.06488} {arXiv:2509.06488 [cond-mat.mtrl-sci]}
  \BibitemShut {NoStop}%
\bibitem [{\citenamefont {Jiang}\ \emph {et~al.}(2017)\citenamefont {Jiang},
  \citenamefont {Thapa}, \citenamefont {Sanders}, \citenamefont {Stanton},
  \citenamefont {Zhang}, \citenamefont {Kono}, \citenamefont {Lou},
  \citenamefont {Chang}, \citenamefont {Hawkins}, \citenamefont {Klem},
  \citenamefont {Pan}, \citenamefont {Smirnov},\ and\ \citenamefont
  {Jiang}}]{PhysRevB.95.045116}%
  \BibitemOpen
  \bibfield  {author} {\bibinfo {author} {\bibfnamefont {Y.}~\bibnamefont
  {Jiang}}, \bibinfo {author} {\bibfnamefont {S.}~\bibnamefont {Thapa}},
  \bibinfo {author} {\bibfnamefont {G.~D.}\ \bibnamefont {Sanders}}, \bibinfo
  {author} {\bibfnamefont {C.~J.}\ \bibnamefont {Stanton}}, \bibinfo {author}
  {\bibfnamefont {Q.}~\bibnamefont {Zhang}}, \bibinfo {author} {\bibfnamefont
  {J.}~\bibnamefont {Kono}}, \bibinfo {author} {\bibfnamefont {W.~K.}\
  \bibnamefont {Lou}}, \bibinfo {author} {\bibfnamefont {K.}~\bibnamefont
  {Chang}}, \bibinfo {author} {\bibfnamefont {S.~D.}\ \bibnamefont {Hawkins}},
  \bibinfo {author} {\bibfnamefont {J.~F.}\ \bibnamefont {Klem}}, \bibinfo
  {author} {\bibfnamefont {W.}~\bibnamefont {Pan}}, \bibinfo {author}
  {\bibfnamefont {D.}~\bibnamefont {Smirnov}}, \ and\ \bibinfo {author}
  {\bibfnamefont {Z.}~\bibnamefont {Jiang}},\ }\href {\doibase
  10.1103/PhysRevB.95.045116} {\bibfield  {journal} {\bibinfo  {journal} {Phys.
  Rev. B}\ }\textbf {\bibinfo {volume} {95}},\ \bibinfo {pages} {045116}
  (\bibinfo {year} {2017})}\BibitemShut {NoStop}%
\bibitem [{\citenamefont {Du}\ \emph {et~al.}(2017)\citenamefont {Du},
  \citenamefont {Li}, \citenamefont {Lou}, \citenamefont {Wu}, \citenamefont
  {Liu}, \citenamefont {Han}, \citenamefont {Zhang}, \citenamefont {Sullivan},
  \citenamefont {Ikhlassi}, \citenamefont {Chang},\ and\ \citenamefont
  {Du}}]{PhysRevLett.119.056803}%
  \BibitemOpen
  \bibfield  {author} {\bibinfo {author} {\bibfnamefont {L.}~\bibnamefont
  {Du}}, \bibinfo {author} {\bibfnamefont {T.}~\bibnamefont {Li}}, \bibinfo
  {author} {\bibfnamefont {W.}~\bibnamefont {Lou}}, \bibinfo {author}
  {\bibfnamefont {X.}~\bibnamefont {Wu}}, \bibinfo {author} {\bibfnamefont
  {X.}~\bibnamefont {Liu}}, \bibinfo {author} {\bibfnamefont {Z.}~\bibnamefont
  {Han}}, \bibinfo {author} {\bibfnamefont {C.}~\bibnamefont {Zhang}}, \bibinfo
  {author} {\bibfnamefont {G.}~\bibnamefont {Sullivan}}, \bibinfo {author}
  {\bibfnamefont {A.}~\bibnamefont {Ikhlassi}}, \bibinfo {author}
  {\bibfnamefont {K.}~\bibnamefont {Chang}}, \ and\ \bibinfo {author}
  {\bibfnamefont {R.-R.}\ \bibnamefont {Du}},\ }\href {\doibase
  10.1103/PhysRevLett.119.056803} {\bibfield  {journal} {\bibinfo  {journal}
  {Phys. Rev. Lett.}\ }\textbf {\bibinfo {volume} {119}},\ \bibinfo {pages}
  {056803} (\bibinfo {year} {2017})}\BibitemShut {NoStop}%
\bibitem [{\citenamefont {Meyer}\ \emph {et~al.}(2025)\citenamefont {Meyer},
  \citenamefont {Baumbach}, \citenamefont {Krishtopenko}, \citenamefont {Wolf},
  \citenamefont {Emmerling}, \citenamefont {Schmid}, \citenamefont {Kamp},
  \citenamefont {Jouault}, \citenamefont {Rodriguez}, \citenamefont {Tournie},
  \citenamefont {Müller}, \citenamefont {Thomale}, \citenamefont {Bastard},
  \citenamefont {Teppe}, \citenamefont {Hartmann},\ and\ \citenamefont
  {Höfling}}]{meyer2025}%
  \BibitemOpen
  \bibfield  {author} {\bibinfo {author} {\bibfnamefont {M.}~\bibnamefont
  {Meyer}}, \bibinfo {author} {\bibfnamefont {J.}~\bibnamefont {Baumbach}},
  \bibinfo {author} {\bibfnamefont {S.}~\bibnamefont {Krishtopenko}}, \bibinfo
  {author} {\bibfnamefont {A.}~\bibnamefont {Wolf}}, \bibinfo {author}
  {\bibfnamefont {M.}~\bibnamefont {Emmerling}}, \bibinfo {author}
  {\bibfnamefont {S.}~\bibnamefont {Schmid}}, \bibinfo {author} {\bibfnamefont
  {M.}~\bibnamefont {Kamp}}, \bibinfo {author} {\bibfnamefont {B.}~\bibnamefont
  {Jouault}}, \bibinfo {author} {\bibfnamefont {J.-B.}\ \bibnamefont
  {Rodriguez}}, \bibinfo {author} {\bibfnamefont {E.}~\bibnamefont {Tournie}},
  \bibinfo {author} {\bibfnamefont {T.}~\bibnamefont {Müller}}, \bibinfo
  {author} {\bibfnamefont {R.}~\bibnamefont {Thomale}}, \bibinfo {author}
  {\bibfnamefont {G.}~\bibnamefont {Bastard}}, \bibinfo {author} {\bibfnamefont
  {F.}~\bibnamefont {Teppe}}, \bibinfo {author} {\bibfnamefont
  {F.}~\bibnamefont {Hartmann}}, \ and\ \bibinfo {author} {\bibfnamefont
  {S.}~\bibnamefont {Höfling}},\ }\href {\doibase 10.1126/sciadv.adz2408}
  {\enquote {\bibinfo {title} {Quantum spin {H}all effect in {III-V}
  semiconductors at elevated temperatures: Advancing topological
  electronics},}\ } (\bibinfo {year} {2025}),\ \Eprint
  {http://arxiv.org/abs/https://www.science.org/doi/pdf/10.1126/sciadv.adz2408}
  {https://www.science.org/doi/pdf/10.1126/sciadv.adz2408} \BibitemShut
  {NoStop}%
\bibitem [{\citenamefont {Hou}\ \emph {et~al.}(2025)\citenamefont {Hou},
  \citenamefont {Wang}, \citenamefont {Jin}, \citenamefont {Zhang},
  \citenamefont {Wang}, \citenamefont {Gong}, \citenamefont {Lian},
  \citenamefont {Shi},\ and\ \citenamefont {Wang}}]{10.1063/5.0237686}%
  \BibitemOpen
  \bibfield  {author} {\bibinfo {author} {\bibfnamefont {Y.-N.}\ \bibnamefont
  {Hou}}, \bibinfo {author} {\bibfnamefont {B.-J.}\ \bibnamefont {Wang}},
  \bibinfo {author} {\bibfnamefont {C.-D.}\ \bibnamefont {Jin}}, \bibinfo
  {author} {\bibfnamefont {H.}~\bibnamefont {Zhang}}, \bibinfo {author}
  {\bibfnamefont {J.-L.}\ \bibnamefont {Wang}}, \bibinfo {author}
  {\bibfnamefont {P.-L.}\ \bibnamefont {Gong}}, \bibinfo {author}
  {\bibfnamefont {R.-Q.}\ \bibnamefont {Lian}}, \bibinfo {author}
  {\bibfnamefont {X.-Q.}\ \bibnamefont {Shi}}, \ and\ \bibinfo {author}
  {\bibfnamefont {R.-N.}\ \bibnamefont {Wang}},\ }\href {\doibase
  10.1063/5.0237686} {\bibfield  {journal} {\bibinfo  {journal} {Journal of
  Applied Physics}\ }\textbf {\bibinfo {volume} {137}},\ \bibinfo {pages}
  {064302} (\bibinfo {year} {2025})},\ \Eprint
  {http://arxiv.org/abs/https://pubs.aip.org/aip/jap/article-pdf/doi/10.1063/5.0237686/20390050/064302\_1\_5.0237686.pdf}
  {https://pubs.aip.org/aip/jap/article-pdf/doi/10.1063/5.0237686/20390050/064302\_1\_5.0237686.pdf}
  \BibitemShut {NoStop}%
\bibitem [{\citenamefont {Jiang}\ \emph {et~al.}(2020)\citenamefont {Jiang},
  \citenamefont {Li}, \citenamefont {Wang}, \citenamefont {Chen}, \citenamefont
  {Chen}, \citenamefont {Xiang}, \citenamefont {Xie}, \citenamefont {Dai},
  \citenamefont {Zhu}, \citenamefont {Yang}, \citenamefont {Sun},\ and\
  \citenamefont {Wen}}]{PhysRevB.101.121115}%
  \BibitemOpen
  \bibfield  {author} {\bibinfo {author} {\bibfnamefont {W.}~\bibnamefont
  {Jiang}}, \bibinfo {author} {\bibfnamefont {B.}~\bibnamefont {Li}}, \bibinfo
  {author} {\bibfnamefont {X.}~\bibnamefont {Wang}}, \bibinfo {author}
  {\bibfnamefont {G.}~\bibnamefont {Chen}}, \bibinfo {author} {\bibfnamefont
  {T.}~\bibnamefont {Chen}}, \bibinfo {author} {\bibfnamefont {Y.}~\bibnamefont
  {Xiang}}, \bibinfo {author} {\bibfnamefont {W.}~\bibnamefont {Xie}}, \bibinfo
  {author} {\bibfnamefont {Y.}~\bibnamefont {Dai}}, \bibinfo {author}
  {\bibfnamefont {X.}~\bibnamefont {Zhu}}, \bibinfo {author} {\bibfnamefont
  {H.}~\bibnamefont {Yang}}, \bibinfo {author} {\bibfnamefont {J.}~\bibnamefont
  {Sun}}, \ and\ \bibinfo {author} {\bibfnamefont {H.-H.}\ \bibnamefont
  {Wen}},\ }\href {\doibase 10.1103/PhysRevB.101.121115} {\bibfield  {journal}
  {\bibinfo  {journal} {Phys. Rev. B}\ }\textbf {\bibinfo {volume} {101}},\
  \bibinfo {pages} {121115} (\bibinfo {year} {2020})}\BibitemShut {NoStop}%
\bibitem [{\citenamefont {Rukelj}\ \emph {et~al.}(2024)\citenamefont {Rukelj},
  \citenamefont {Željana Bonačić~Lošić}, \citenamefont {Kupčić},\ and\
  \citenamefont {Akrap}}]{Rukelj}%
  \BibitemOpen
  \bibfield  {author} {\bibinfo {author} {\bibfnamefont {Z.}~\bibnamefont
  {Rukelj}}, \bibinfo {author} {\bibnamefont {Željana Bonačić~Lošić}},
  \bibinfo {author} {\bibfnamefont {I.}~\bibnamefont {Kupčić}}, \ and\
  \bibinfo {author} {\bibfnamefont {A.}~\bibnamefont {Akrap}},\ }\href
  {\doibase https://doi.org/10.1016/j.physb.2024.416590} {\bibfield  {journal}
  {\bibinfo  {journal} {Physica B: Condensed Matter}\ }\textbf {\bibinfo
  {volume} {695}},\ \bibinfo {pages} {416590} (\bibinfo {year}
  {2024})}\BibitemShut {NoStop}%
\bibitem [{\citenamefont {Zhang}\ \emph {et~al.}(2026)\citenamefont {Zhang},
  \citenamefont {Jia}, \citenamefont {kai Lou}, \citenamefont {Wang},
  \citenamefont {Su}, \citenamefont {Chang},\ and\ \citenamefont {Du}}]{Zhang}%
  \BibitemOpen
  \bibfield  {author} {\bibinfo {author} {\bibfnamefont {W.}~\bibnamefont
  {Zhang}}, \bibinfo {author} {\bibfnamefont {P.}~\bibnamefont {Jia}}, \bibinfo
  {author} {\bibfnamefont {W.}~\bibnamefont {kai Lou}}, \bibinfo {author}
  {\bibfnamefont {X.}~\bibnamefont {Wang}}, \bibinfo {author} {\bibfnamefont
  {S.}~\bibnamefont {Su}}, \bibinfo {author} {\bibfnamefont {K.}~\bibnamefont
  {Chang}}, \ and\ \bibinfo {author} {\bibfnamefont {R.-R.}\ \bibnamefont
  {Du}},\ }\href {\doibase 10.1088/0256-307X/43/1/010704} {\bibfield  {journal}
  {\bibinfo  {journal} {Chin. Phys. Lett.}\ }\textbf {\bibinfo {volume} {43}},\
  \bibinfo {pages} {010704} (\bibinfo {year} {2026})}\BibitemShut {NoStop}%
\bibitem [{\citenamefont {Sablikov}(2025)}]{Sablikov1}%
  \BibitemOpen
  \bibfield  {author} {\bibinfo {author} {\bibfnamefont {V.~A.}\ \bibnamefont
  {Sablikov}},\ }\href {\doibase https://doi.org/10.1016/j.physe.2025.116213}
  {\bibfield  {journal} {\bibinfo  {journal} {Physica E: Low-dimensional
  Systems and Nanostructures}\ }\textbf {\bibinfo {volume} {170}},\ \bibinfo
  {pages} {116213} (\bibinfo {year} {2025})}\BibitemShut {NoStop}%
\bibitem [{\citenamefont {Das}\ \emph {et~al.}(2019)\citenamefont {Das},
  \citenamefont {Wickramaratne}, \citenamefont {Debnath}, \citenamefont {Yin},\
  and\ \citenamefont {Lake}}]{PhysRevB.99.085409}%
  \BibitemOpen
  \bibfield  {author} {\bibinfo {author} {\bibfnamefont {P.}~\bibnamefont
  {Das}}, \bibinfo {author} {\bibfnamefont {D.}~\bibnamefont {Wickramaratne}},
  \bibinfo {author} {\bibfnamefont {B.}~\bibnamefont {Debnath}}, \bibinfo
  {author} {\bibfnamefont {G.}~\bibnamefont {Yin}}, \ and\ \bibinfo {author}
  {\bibfnamefont {R.~K.}\ \bibnamefont {Lake}},\ }\href {\doibase
  10.1103/PhysRevB.99.085409} {\bibfield  {journal} {\bibinfo  {journal} {Phys.
  Rev. B}\ }\textbf {\bibinfo {volume} {99}},\ \bibinfo {pages} {085409}
  (\bibinfo {year} {2019})}\BibitemShut {NoStop}%
\bibitem [{\citenamefont {Shchamkhalova}\ and\ \citenamefont
  {Sablikov}(2025)}]{Shchamkhalova}%
  \BibitemOpen
  \bibfield  {author} {\bibinfo {author} {\bibfnamefont {B.~S.}\ \bibnamefont
  {Shchamkhalova}}\ and\ \bibinfo {author} {\bibfnamefont {V.~A.}\ \bibnamefont
  {Sablikov}},\ }\href {\doibase https://doi.org/10.1016/j.physb.2025.417942}
  {\bibfield  {journal} {\bibinfo  {journal} {Physica B: Condensed Matter}\
  }\textbf {\bibinfo {volume} {719}},\ \bibinfo {pages} {417942} (\bibinfo
  {year} {2025})}\BibitemShut {NoStop}%
\bibitem [{\citenamefont {Sablikov}\ and\ \citenamefont
  {Sukhanov}(2023{\natexlab{a}})}]{SABLIKOV2023115492}%
  \BibitemOpen
  \bibfield  {author} {\bibinfo {author} {\bibfnamefont {V.~A.}\ \bibnamefont
  {Sablikov}}\ and\ \bibinfo {author} {\bibfnamefont {A.~A.}\ \bibnamefont
  {Sukhanov}},\ }\href {\doibase https://doi.org/10.1016/j.physe.2022.115492}
  {\bibfield  {journal} {\bibinfo  {journal} {Physica E: Low-dim.Systems and
  Nanostructures}\ }\textbf {\bibinfo {volume} {145}},\ \bibinfo {pages}
  {115492} (\bibinfo {year} {2023}{\natexlab{a}})}\BibitemShut {NoStop}%
\bibitem [{\citenamefont {Sablikov}\ and\ \citenamefont
  {Sukhanov}(2023{\natexlab{b}})}]{SABLIKOV2023129006}%
  \BibitemOpen
  \bibfield  {author} {\bibinfo {author} {\bibfnamefont {V.~A.}\ \bibnamefont
  {Sablikov}}\ and\ \bibinfo {author} {\bibfnamefont {A.~A.}\ \bibnamefont
  {Sukhanov}},\ }\href {\doibase
  https://doi.org/10.1016/j.physleta.2023.129006} {\bibfield  {journal}
  {\bibinfo  {journal} {Physics Letters A}\ }\textbf {\bibinfo {volume}
  {481}},\ \bibinfo {pages} {129006} (\bibinfo {year}
  {2023}{\natexlab{b}})}\BibitemShut {NoStop}%
\bibitem [{\citenamefont {Bernevig}\ \emph {et~al.}(2006)\citenamefont
  {Bernevig}, \citenamefont {Hughes},\ and\ \citenamefont {Zhang}}]{BHZ}%
  \BibitemOpen
  \bibfield  {author} {\bibinfo {author} {\bibfnamefont {B.~A.}\ \bibnamefont
  {Bernevig}}, \bibinfo {author} {\bibfnamefont {T.~L.}\ \bibnamefont
  {Hughes}}, \ and\ \bibinfo {author} {\bibfnamefont {S.-C.}\ \bibnamefont
  {Zhang}},\ }\href {\doibase 10.1126/science.1133734} {\bibfield  {journal}
  {\bibinfo  {journal} {Science}\ }\textbf {\bibinfo {volume} {314}},\ \bibinfo
  {pages} {1757} (\bibinfo {year} {2006})}\BibitemShut {NoStop}%
\bibitem [{\citenamefont {K\"{o}nig}\ \emph {et~al.}(2008)\citenamefont
  {K\"{o}nig}, \citenamefont {Buhmann}, \citenamefont {W.~Molenkamp},
  \citenamefont {Hughes}, \citenamefont {Liu}, \citenamefont {Qi},\ and\
  \citenamefont {Zhang}}]{JPSJ.77.031007}%
  \BibitemOpen
  \bibfield  {author} {\bibinfo {author} {\bibfnamefont {M.}~\bibnamefont
  {K\"{o}nig}}, \bibinfo {author} {\bibfnamefont {H.}~\bibnamefont {Buhmann}},
  \bibinfo {author} {\bibfnamefont {L.}~\bibnamefont {W.~Molenkamp}}, \bibinfo
  {author} {\bibfnamefont {T.}~\bibnamefont {Hughes}}, \bibinfo {author}
  {\bibfnamefont {C.-X.}\ \bibnamefont {Liu}}, \bibinfo {author} {\bibfnamefont
  {X.-L.}\ \bibnamefont {Qi}}, \ and\ \bibinfo {author} {\bibfnamefont {S.-C.}\
  \bibnamefont {Zhang}},\ }\href {\doibase 10.1143/JPSJ.77.031007} {\bibfield
  {journal} {\bibinfo  {journal} {J. Phys. Society of Japan}\ }\textbf
  {\bibinfo {volume} {77}},\ \bibinfo {pages} {031007} (\bibinfo {year}
  {2008})},\ \Eprint
  {http://arxiv.org/abs/https://doi.org/10.1143/JPSJ.77.031007}
  {https://doi.org/10.1143/JPSJ.77.031007} \BibitemShut {NoStop}%
\bibitem [{\citenamefont {Schliemann}\ and\ \citenamefont {Loss}(2003)}]{Loss}%
  \BibitemOpen
  \bibfield  {author} {\bibinfo {author} {\bibfnamefont {J.}~\bibnamefont
  {Schliemann}}\ and\ \bibinfo {author} {\bibfnamefont {D.}~\bibnamefont
  {Loss}},\ }\href {\doibase 10.1103/PhysRevB.68.165311} {\bibfield  {journal}
  {\bibinfo  {journal} {Phys. Rev. B}\ }\textbf {\bibinfo {volume} {68}},\
  \bibinfo {pages} {165311} (\bibinfo {year} {2003})}\BibitemShut {NoStop}%
\bibitem [{\citenamefont {Culcer}(2012)}]{CULCER2012860}%
  \BibitemOpen
  \bibfield  {author} {\bibinfo {author} {\bibfnamefont {D.}~\bibnamefont
  {Culcer}},\ }\href {\doibase https://doi.org/10.1016/j.physe.2011.11.003}
  {\bibfield  {journal} {\bibinfo  {journal} {Physica E: Low-dimensional
  Systems and Nanostructures}\ }\textbf {\bibinfo {volume} {44}},\ \bibinfo
  {pages} {860} (\bibinfo {year} {2012})},\ \bibinfo {note} {sI:Topological
  Insulators}\BibitemShut {NoStop}%
\bibitem [{\citenamefont {Bandyopadhyay}\ \emph {et~al.}(2024)\citenamefont
  {Bandyopadhyay}, \citenamefont {Joseph},\ and\ \citenamefont
  {Narayan}}]{BANDYOPAD}%
  \BibitemOpen
  \bibfield  {author} {\bibinfo {author} {\bibfnamefont {A.}~\bibnamefont
  {Bandyopadhyay}}, \bibinfo {author} {\bibfnamefont {N.~B.}\ \bibnamefont
  {Joseph}}, \ and\ \bibinfo {author} {\bibfnamefont {A.}~\bibnamefont
  {Narayan}},\ }\href {\doibase https://doi.org/10.1016/j.mtelec.2024.100101}
  {\bibfield  {journal} {\bibinfo  {journal} {Materials Today Electronics}\
  }\textbf {\bibinfo {volume} {8}},\ \bibinfo {pages} {100101} (\bibinfo {year}
  {2024})}\BibitemShut {NoStop}%
\end{thebibliography}
merlin.mbs apsrev4-1.bst 2010-07-25 4.21a (PWD, AO, DPC) hacked

\end{document}